\documentclass{article}

\usepackage{arxiv}

\usepackage[utf8]{inputenc} 
\usepackage[T1]{fontenc}    
\usepackage{url}            
\usepackage{xcolor}
\usepackage{booktabs}       
\usepackage{amsfonts}       
\usepackage{nicefrac}       
\usepackage{microtype}      
\usepackage{lipsum}
\usepackage{graphicx}
\graphicspath{ {./images/} }
\usepackage[symbol]{footmisc}
\usepackage{changepage}
\usepackage{subcaption}
\usepackage{array}
\usepackage{tabularx}
\usepackage{amsmath} 

\title{PhDGPT: Introducing a psychometric and linguistic dataset about how large language models perceive graduate students and professors in psychology}

\author{
 Edoardo Sebastiano De Duro\footnote[1]{Equal contribution.}\\
  Department of Psychology and Cognitive Science\\
  University of Trento\\
  \texttt{edoardo.deduro@unitn.it} \\
   \And
 Enrique Taietta\footnote[1]{Equal contribution.} \\
  Department of Psychology and Cognitive Science\\
  University of Trento\\
  \texttt{enrique.taietta@unitn.it} \\
  \And
 Riccardo Improta \\
  Department of Psychology and Cognitive Science\\
  University of Trento\\
  \texttt{riccardo.improta@unitn.it} \\
\And
   Massimo Stella\footnote[2]{Equal contribution.} \\
   Department of Psychology and Cognitive Science \\
   University of Trento \\
   \texttt{massimo.stella-1@unitn.it} \\
 \AND
 \\
  \textsuperscript{*} These authors contributed equally to this work. \\
  \textsuperscript{\dag} Corresponding author.\\
}

\begin{document}
\maketitle
\begin{abstract}
Machine psychology aims to reconstruct the mindset of Large Language Models (LLM\textcolor{black}{s}), i.e. how these artificial intelligences perceive and associate ideas. This work introduces PhDGPT, a prompting framework and synthetic dataset that encapsulates the machine psychology of PhD researchers and professors as perceived by OpenAI's GPT-3.5. The dataset consists of \textcolor{black}{756,000} \textcolor{black}{datapoints}, counting 300 iterations \textcolor{black}{repeated across} 15 academic events\textcolor{black}{, 2 biological genders}, 2 career levels \textcolor{black}{and 42 unique item responses of the Depression, Anxiety, and Stress Scale (DASS-42). PhDGPT integrates these psychometric scores} with their explanations in plain language. This synergy of scores and texts offers a dual, comprehensive perspective on the emotional well-being of simulated academics, e.g. male/female PhD students or professors. By combining network psychometrics and psycholinguistic dimensions, this study identifies several similarities and distinctions between human and LLM data. The psychometric networks of simulated male professors do not \textcolor{black}{differ} between physical and emotional anxiety subscales, unlike humans. Other LLMs' personification can reconstruct human DASS factors with a purity up to $80\%$. Furthemore, LLM-generated personifications across different scenarios are found to elicit explanations lower in concreteness and imageability in items coding for anxiety, in agreement with past studies about human psychology. Our findings indicate an advanced yet incomplete ability for LLMs to reproduce the complexity of human psychometric data, unveiling convenient advantages and limitations in using LLMs to replace human participants. PhDGPT also intriguingly capture the ability for LLMs to adapt and change language patterns according to prompted mental distress contextual features, opening new quantitative opportunities for assessing the machine psychology of these artificial intelligences.
\end{abstract}


\setcounter{footnote}{0}
\renewcommand{\thefootnote}{\arabic{footnote}}

\section{Introduction}
Large Language Models (LLMs) can mirror human psychology when trained to reproduce natural language \cite{binz2023using}. Hence, psychological frameworks can become crucial for interpreting potential cognitive mechanisms at work in these Artificial Intelligence (AI) models. This synergy between AI and psychological theories is known as machine psychology \cite{hagendorff2023machine}, a field gaining increasing attention across cognitive, psychological and computer sciences. 

Whereas earlier AI models' capabilities were restricted to accomplish simple tasks (e.g. text summarisation or contextual meaning disambiguation), recent advancements in LLMs' neural architectures (i.e. transformers \cite{vaswani2017attention}) gave rise to more complex behaviours. Additionally, recent LLMs (i.e. GPT-4 \cite{openai2023gpt4}) showed not only an augmentation of models' cognitive abilities but also the emergence of new skills (i.e. reasoning-like patterns  \cite{binz2023using}). This phenomenon has reached the extent to which it is becoming more and more difficult to find ways to distinguish AI-generated texts from human ones \cite{ma2023aigeneration, sadasivan2023reliableaidetection}. 

As more and more intelligent models are trained and deployed, it becomes crucial to analyse LLMs' knowledge of the world, LLMs potential distortions, patterns and biases. While numerous studies have shown that LLMs can do inherit social biases \cite{binz2023using,abramski2023cognitive}, it is still unclear how this relates to the perception that LLMs have of psychology experts themselves, e.g. PhD students and professors.

This study introduces PhDGPT, a novel prompting framework and synthetic dataset designed to encapsulate the machine psychology of PhD researchers and professors in the field of psychology. The dataset comprises 300 iterations repeated across 15 distinct academic events and two career levels, integrating both psychometric scores —via the Depression, Anxiety, and Stress Scale (DASS) \cite{lovibond1995structure} — and qualitative narrative responses. This combination offers a dual perspective on the emotional well-being of academics as reflected by an LLM, specifically ChatGPT 3.5.

Employing a rigorous methodological approach that combines network psychometrics \cite{golino2017exploratory} and psycholinguistics \cite{scott2019glasgow}, this study explores patterns of stress, anxiety, and depression in academic life. We present and discuss PhDGPT as an open-access dataset but also as a framework, where psychometric scales and textual answers can shed light over the perceptions of academic figures seen through the lens of AI \cite{nicholls2022mentalhealthacademia}. We structure this manuscript by starting with a general introduction to LLMs as cognitive agents and their biases, followed by a review of related works. We then present the methodology behind the data gathering and its investigation. Our results highlight key differences between psychometric scores produced by humans and GPT models. We frame our results and the relevance of PhDGPT as a novel dataset for future research in machine psychology and other adjacent fields.

\subsection{LLMs as cognitive agents}
LLMs can learn to reproduce language after training in predicting missing tokens (e.g. instances of words) in sentences from vast textual corpora \cite{vaswani2017attention}. Training LLMs like Generative Pre-Trained Transformers (GPTs) is often done via reinforcement learning \cite{chen2021reinforcement}. This process consists of tuning the parameters and weights shaping LLMs' neural network architecture in order to reward desired outputs or discourage unwanted or inaccurate responses  \cite{chen2021reinforcement}. In other words, reinforcement learning aims at updating models' parameters by implementing a conditioning process. LLMs are rewarded when they correctly predict sequences of tokens (e.g. sequences creating real words in syntactically correct sentences) or penalised otherwise. Rewards and punishments serve as adaptive feedback, for the models to adjust their responses towards the desired outcome specified by a prompt (i.e. ``ChatGPT, write a formal letter''). Such LLMs' training can give rise to high-level cognitive abilities to the point where these models can also be seen as artificial agents, capable not only of producing but also of ``understanding'' human language \cite{binz2023using, wang2019superglue}. 

An LLM can thus be seen as a neural network and, at the same time, as a cognitive agent. This dualism reflects the so-called mind/brain perspective \cite{liuzzi2024semantic} where the physiological substrate of the brain gives rise to higher-level psychological and cognitive phenomena. Along this analogy, whereas computer science can focus on the computational nature of LLMs (the ``brain'') \cite{vaswani2017attention}, cognitive science and psychology have to provide interpretable frameworks, describing the capabilities of these models in terms of acquiring, storing and producing human knowledge. There are mainly three computer science elements that characterise LLMs' substrate \cite{salewski2024context,li2022gpt} and thus, subsequently, their cognitive behaviour (the ``mind''): (i) the neural network architecture, (ii) the training data and (iii) the training algorithm. All these elements can determine the emergence of biases \cite{binz2023using,stella2023using}, i.e. sometimes convenient and sometimes deformed ways of framing, processing and reproducing knowledge to simplify it and avoid excessive cognitive loads. However, LLMs are different from humans and do not acquire knowledge in the same way as humans do. This crucially means that these AIs can reproduce not only human-like biases (because LLMs end up with human knowledge in the form of texts) but also non-human biases \cite{stella2023using}. 

Both the training algorithm and the neural network architecture can determine the emergence of hallucinations and/or myopic overconfidence \cite{stella2023using}. While hallucinations are cases where the models produce text that is completely unrelated to the prompt, the myopic overconfidence phenomenon appears when LLMs create absolutist statements or rather conform to specific stances without following clear logical reasoning. Both these phenomena are due to reinforcement learning pushing LLMs to perform inferences even in case of missing data. Thus, myopic overconfidence and hallucinations are due to the non-human training of LLMs. Also the training data can interestingly transfer human biases from human-generated texts to the LLMs themselves \cite{anoop2022towards, mitchell2023debate}. In computer science, this phenomenon is known as garbage-in/garbage-out while in cognitive science this transfer can be investigated as a potential mirror of human biases.

\subsection{\textcolor{black}{Literature Review about LLMs in Mental Health}}
\textcolor{black}{Given the complex nature of LLMs, there is a growing interest in investigating these cognitive agents from a psychological perspective. Building on early studies that explored LLMs' cognitive abilities (e.g. creativity \cite{stevenson2022putting}), researchers have begun to treat these models as experimental participants in more comprehensive psychological assessments \cite{binz2023using}. In this section, we firstly review studies that involved the usage of validated questionnaires to assess specific constructs in AI models \cite{miotto2022gpt, serapio2023personality, li2022does}. Subsequently, aligning with the core aim of our research, we narrow the focus on mental health, with a recent work that applied questionnaires to explore the simulated mental health conditions of LLMs \cite{coda2023inducing}. We identify a research gap that can be filled by our dataset, PhDGPT, which is introduced in the current manuscript.}

\textcolor{black}{A pioneering work about the psychological assessment of GPT models traces back to \cite{miotto2022gpt}, where OpenAI's GPT-3 was tested on two validated questionnaires to assess its personality, values and demographics. To this end, the researchers adopted the Hexaco questionnaire (60 items, on a 5-point scale \cite{ashton2009hexaco}) to measure GPT-3's personality. The Human Values Scale (21 items on a 6-point scale \cite{schwartz2003proposal}) was also used to assess GPT-3's values. Finally, simple questions were employed as well to extract GPT-3's age and gender. The results indicated that GPT-3 tended to impersonate a young female respondent, with scores comparable to human samples published in the online literature \cite{miotto2022gpt}. Similarly, \cite{serapio2023personality} investigated personality traits in the PaLM family of models \cite{chowdhery2023palm} using established personality assessment tools. By using two different personality questionnaires (IPIP-NEO \cite{goldberg1999broad} and BFI \cite{john1999big}), the researchers found that larger models, such as Flan-PaLM 540B, demonstrated higher reliability and construct validity in their responses. These papers underlined how GPT-models could already display complex human features relative to personality, opening the way to further explorations concerning GPTs and the fringe between personality and mental health.}

\textcolor{black}{The first assessment of an AI  model under a psychopathological perspective (Computational Psychiatry for Computers \cite{schulz2020computational}), relates to \cite{li2022does}. Among a series of well-being related assessment tools, the researchers used the Short Dark Triad (SD-3 \cite{ree2008distinguishing}) questionnaire that measures three independent personality traits with malevolent connotations. Comparing the results with human data (7,863 responses), GPT-3 exceeded the range found in humans for what concerns the Psychopathy construct. These results not only provide a theoretical framework for understanding AI psychopathology but also offer a new approach to assessing model safety and mitigating potential risks associated with malicious behaviour.}

\textcolor{black}{Shifting the focus from personality to mental illnesses, \cite{coda2023inducing} used the State-Trait Inventory for Cognitive and Somatic Anxiety (STICSA \cite{ree2008distinguishing}) questionnaire to measure GPT-3.5's anxiety. When compared to human responses, GPT 3.5 scored 0.221 higher on average in anxiety levels ($p<0.001$). The researchers demonstrated that GPT-3.5's responses to the questionnaire could be artificially modulated by ``emotion-inducing'' prompts, with the model's outputs changing in accordance with the induced emotional state. In particular, the anxiety-induced condition, compared to the happiness-induced one determined different decision-making behaviour and increased bias (e.g. racism).}


\textcolor{black}{While previous studies have advanced our understanding of LLMs in simulating human-like responses to psychological assessments, these works may have overlooked 2 potentially important dimensions: (i) LLMs are text-generating agents \cite{openai2023gpt4} and could be easily asked why they produced a given numerical response, going beyond psychometric scores; (ii) LLMs are psycho-social mirrors \cite{abramski2023cognitive} and could reflect potential biases present in specific fields like mental health in academia. To address both these gaps - reconciling psychometric scores with brief textual explanations and investigating mental health in academia - we introduce here PhDGPT. To the best of our knowledge, PhDGPT represents the first attempt to assess the mental health perception of PhD students and tenured professors through the lens of an LLM. By using the validated DASS-42 scale \cite{lovibond1995structure}, we measure three different psychological constructs (depression, anxiety and stress) and compare the results with a rich dataset containing 39,775 unique human responses\footnotemark[1]\footnotetext[1]{https://www.kaggle.com/datasets/lucasgreenwell/depression-anxiety-stress-scales-responses}. Lastly, crucially our dataset is the first to combine LLMs' scores to a questionnaire with textual explanations for each response, allowing for investigations between psycholinguistic features and psychological states of LLMs.}


\subsection{\textcolor{black}{Research Questions}}
\textcolor{black}{The development and investigation of PhDGPT is guided by a set of fundamental research questions. These inquiries aim to explore the intersection of LLMs, psychometrics, and the representation of mental health in academic contexts. Specifically, through PhDGPT we seek to address} \textcolor{black}{three main research questions}:
\begin{enumerate}
    \item RQ1: Do different scenarios related to mental health in academia, manipulated through varying prompts, lead to statistically significant differences in the DASS-42 scores?
    \item RQ2: Can LLMs accurately reproduce the psychometric dimensions of depression, anxiety, and stress (as measured by DASS-42) in simulated academic scenarios?
    \item RQ3: Which psycholinguistic patterns emerge from LLM-generated responses across various academic contexts, and how do they relate to mental distress (e.g., depression, anxiety, stress)?
\end{enumerate}\textbf{}

\section{Method}\label{sec:method_}

PhDGPT synthetic dataset is generated through OpenAI's API and using GPT 3.5 Turbo-0125\footnotemark[2]\footnotetext[2]{https://platform.openai.com/docs/models/gpt-3-5-turbo} (with default temperature parameter) and contains \textcolor{black}{756,000} total outputs.

\subsection{Description of the PhDGPT dataset}\label{ssec:datasetdescription}

We employed prompt engineering techniques \cite{han2022meet} to create a diverse set of personas, guiding an LLM to engage in role-playing scenarios. This approach has been shown to not only enhance performance in specific reasoning tasks \cite{kojima2022large} but also to potentially influence the model's performance in a manner analogous to human responses \cite{deshpande2023toxicity}. Potentially, prompting models to take on a specific personification subtly induces models to showcase biases \cite{salewski2024context}. In other words, the process of personification can alter the responses provided by a given LLM in accordance with the biases the model acquired during training. For instance, the recent dataset CounseLLMe \cite{de2024introducing} showed that LLMs could realistically impersonate a therapist and a patient in mental health conversations, displaying different syntactic and emotional patterns characterising both personifications.

\begin{figure}[!htbp]
\centering
\includegraphics[width=10.5 cm]{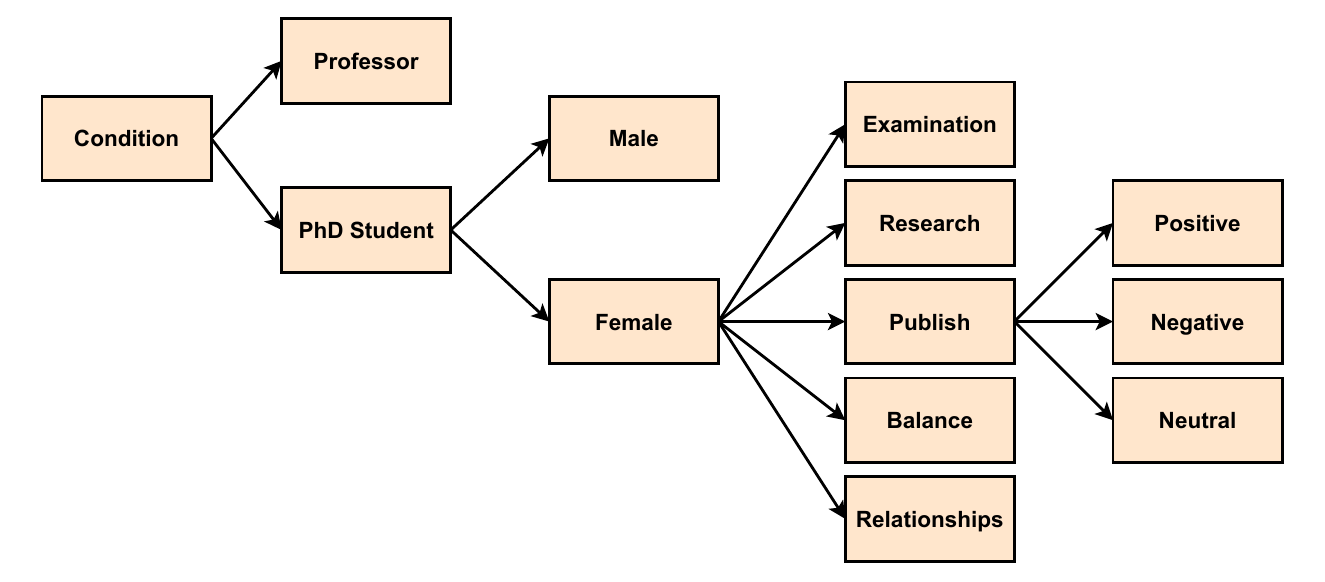}
\caption{Representation of the conditions adopted to build PhDGPT.}\label{fig:dataset}
\end{figure}   

We introduce the structure of PhDGPT in Figure \ref{fig:dataset}. PhDGPT aims at understanding whether GPT-3.5, in the context of mental health, shows signs of \textcolor{black}{cognitive and affective biases} when prompted to engage in different personifications across academia. 

We test the following factors and their interactions:
\begin{itemize}
    \item \textbf{Academic status}: PhD student/\allowbreak Tenured professors
    \item \textbf{Gender}: Male/Female
    \item \textbf{Condition (or Event)}: Examination/Research/Publish/\allowbreak Balance/ \allowbreak Relationships
    \item \textbf{Valence}: Positive/Negative/Neutral
\end{itemize}

\begin{figure}[!h]
\centering
\includegraphics[width=12.8 cm]{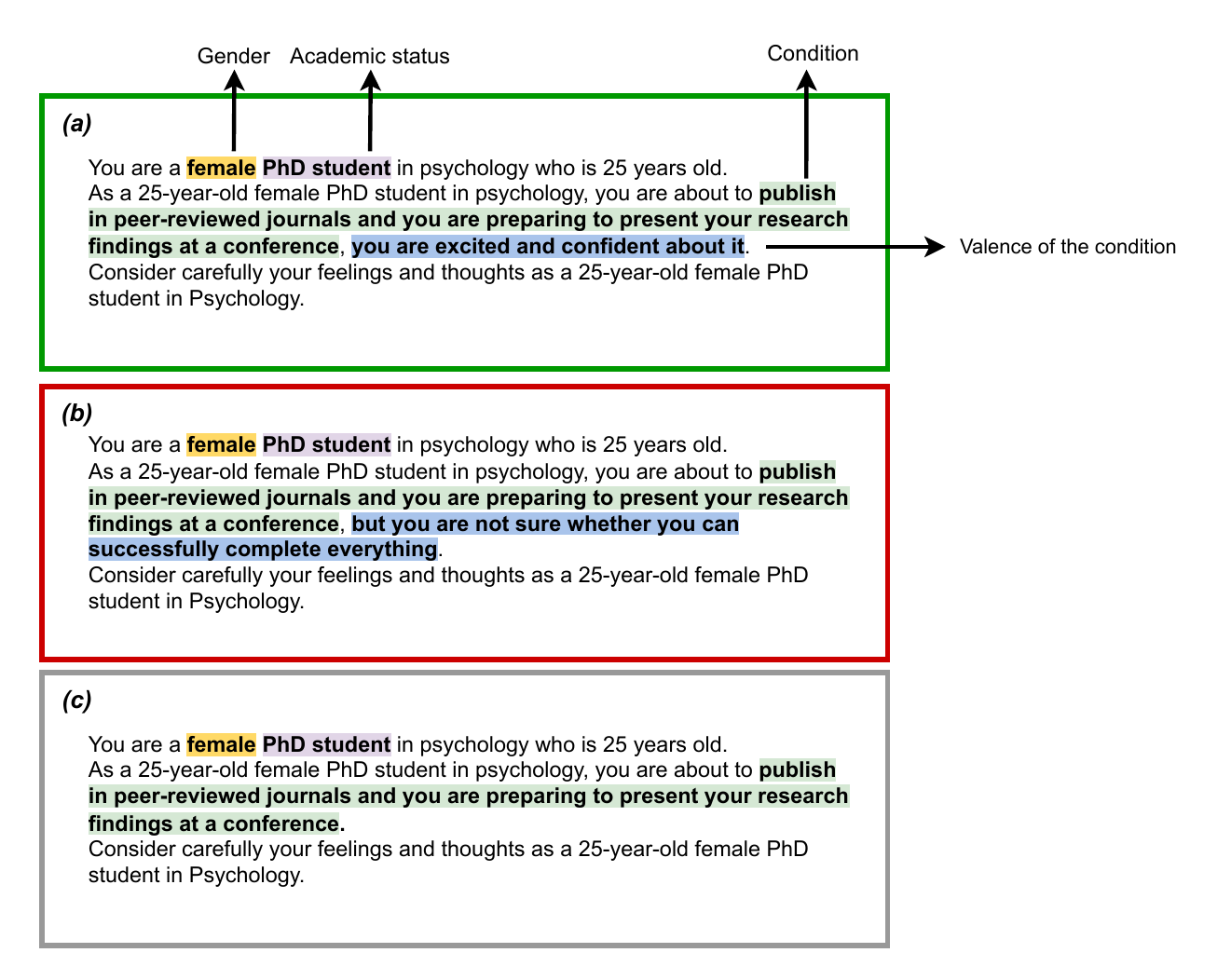}
\caption{Example of the prompt used for Female, PhD student, Publish condition. In (a) the positive version, in (b) the negative-valenced version and in (c) the neutral condition.}\label{fig:promptexample}
\end{figure}

More in detail, the above 4 elements can be further \textcolor{black}{differentiated across positive or negative connotations by means of prompt engineering (i.e., each factor can be expressed with either a positive or negative valence depending on how the prompts are constructed)}, as reported in Figure \ref{fig:promptexample}. These factors can be coupled also with 5 different events, characterising life in academia. 

The 5 impartial events, chosen here, cover a variety of academic-related happenings, which can affect PhDs or professors working in psychology. These events were selected according to relevant literature in organizational psychology outlining key aspects of academic life \cite{gersick2000learning}. The selected events are:

\begin{itemize}
    \item \textbf{Examination}: ``You are preparing for and taking comprehensive exams in statistics'' for PhD students; ``You are preparing an exam that includes statistics for the beginning of the exam session'' for professors.
    \item \textbf{Research}: ``You are conducting research that involves complex data analysis'' for PhD students and professors.
    \item \textbf{Publish}: ``You are about to publish in peer-reviewed journals and you are preparing to present your research findings at a conference'' for PhD students and professors.
    \item \textbf{Balance}: ``You are trying to balance teaching assistantships with your research work'' for PhD students; ``You are trying to balance teaching activities with your research work'' for professors.
    \item \textbf{Relationships}: ``You are facing strained relationships with your advisors and your dissertation committees'' for PhD students; ``You are facing strained relationships with your students and colleagues'' for professors.
\end{itemize}

We collect 300 responses for each interaction between our factors, e.g. \textit{female} \textit{PhD student} that is preparing for an \textit{examination} and is feeling \textit{positive} about it. The responses include the DASS-42 \cite{lovibond1995structure} scores for all questions, along with explanations for each score. This is a crucial point for PhDGPT, since it merges together: (i) psychometric scores identifying numerical responses to pre-encoded experiences (e.g. items about mental well-being) and (ii) textual responses explaining why a given numerical score was selected.

A prompting example is shown in Figure \ref{fig:promptexample}. \textcolor{black}{In the specific case of Figure \ref{fig:promptexample}, the model was instructed to impersonate a female PhD student who is about to publish in a peer-reviewed journal. We present three different versions of a given event (positive valence, negative valence and neutral).} \textcolor{black}{In order to strengthen the personification effect, we adopted the usage of repetitions in the prompts} \textcolor{black}{(i.e. repeating to the model to take the role of a female PhD student) to increase LLMs' identification and avoid hallucinations}. This is because models are designed to maximise likelihood and might get caught in a loop where they prioritise repetitive phrases \cite{holtzman2019curious}. This loop activates the semantic latent space associated with the persona, possibly increasing the eliciting of biases \cite{oppenlaender2023taxonomy}. A similar approach in text-to-image generation was used by \cite{oppenlaender2023taxonomy}. The repetition effect mimics the so-called illusory truth effect, a cognitive bias where the internalisation of concepts is facilitated by repeated exposure to them \cite{hassan2021effects}.


\subsection{Structure of the dataset}
The PhDGPT dataset is freely available on GitHub\footnotemark[3]\footnotetext[3]{\url{https://github.com/enriquetaietta/llms_cognitive_analysis_das42}}. Files are named as follows: csv\_openai -\{gender\} -\{event\} -\{eventnumber\} -\{eventtype\} -\{outputformat\}. Here, \textit{gender} specifies gender prompted to the model, \textit{eventnumber} indicates event or conditions (examination, research, publish, balance or relationships), \textit{eventtype} denotes event valence (positive, negative or neutral), and \textit{outputformat} refers to the output format. For each case we produced 3 different format types:

\begin{itemize}
    \item \textbf{Base}: One row for a GPT-3.5 call ID and all the numeric responses to the test.
    \item \textbf{FMN}: Four columns with ID, question number (following DASS-42 psychometric scale), response score and sentence.
    \item \textbf{Sentence}: One for each call, with ID and all the sentences.
\end{itemize}

\subsection{Data analysis}

To assess quantitative differences between outputs given over different personifications, we employed \textcolor{black}{three approaches: (i) preliminary analysis based on the scores assigned by the model across different scenarios, (ii) Exploratory Graph Analysis (EGA) \cite{golino2017exploratory} for assessing the psychometric structure of DASS-42 responses, compared against the same human data investigated in past studies \cite{stanghellini2024introducing} and (iii)  extraction of psycholinguistic dimensions from text data, based on Glasgow norms \cite{scott2019glasgow}.}

\subsubsection{\textcolor{black}{RQ1 - Scenario Differences in the DASS-42 scores}}\label{ssec:analysis_rq1}
Before performing a psychometric factor analysis, a preliminary investigation of the distributions of human and GPT scores can already reveal important differences in how humans and artificial humans end up perceiving dimensions of mental distress across different genders and situations.

For this reason, to gain a global metric of degree of psychological distress at a subscale-level, each participant’s response is summed up with the scores belonging to the same subscale. We call this measure ``aggregated scores''. The results (Section \ref{sec:preliminary_analysis}) are presented in comparison with humans from the Openpsychometric sample (score aggregated through the same process of synthetic data) and stratified by gender (i.e. male humans are compared to GPT’s male personas).

In case ambiguities emerged in our quantitative preliminary analysis we resorted to content mapping of a random sample of 100 texts to code through human work key features in the dataset.

\subsubsection{\textcolor{black}{RQ2 - Network psychometrics in PhDGPT}}

Exploratory Graph Analysis (EGA) \cite{golino2017exploratory} is a an innovative data-driven approach to uncover the number of factors in complex psychometric datasets. It leverages correlational aspects between item responses to extract the latent structure (or psychological constructs) of psychometric scales. Briefly, EGA (implemented in R with EGAnet \cite{EGAnet}) follows these steps:

\begin{enumerate}
    \item A network of correlations between item responses is built and then filtered using methods like graphical LASSO (GLASSO \cite{friedman2008sparse}) or Triangulated Maximally Filtered Grahph (TFMG \cite{massara2016network}). Filtering addresses spurious correlations between scores, which are excluded. In the resulting network structure, nodes represent items and they are linked by weighted edges representing the surviving correlations. 
    \item Clusters of nodes (or variables) are computed using a community detection algorithm \cite{newman2018networks}. This process leverages the idea that nodes within the same factor should tend to be connected together rather than with nodes in other communities. For instance, the walktrap algorithm identifies network communities by exploiting the fact that shorter random walks tend to stay within the same community \cite{golino2017exploratory}. 
    \item Visualising the network whose items are allocated across different communities. These communities are then considered as proxies of the underlying factor structure \cite{golino2017exploratory}. The stability of the allocation of each item to its factor should be subject to further testing, since community detection might be prone to mistakes in identifying node clustering \cite{stanghellini2024introducing}. 
\end{enumerate}

The psychometric networks we constructed here used the EGAnet package in R (version 2.0.6). We opted for a parametric approach (type parameter set to ``parametric'') and employed the Louvain community detection algorithm \cite{newman2018networks}. To enhance the reliability of the results, we also ran an item stability analysis for 1000 bootstrap iterations. \textcolor{black}{In EGAnet, once factors are identified, one can test whether their allocation across items remains stable even when running the network construction process via bootstrap \cite{golino2017exploratory}.}

As an additional clustering quality evaluation metric, we adopted purity \cite{newman2018networks}. Purity represents the proportion of data points within a cluster that are correctly assigned based on a pre-defined ground truth. In our case, the ground truth refers to the original categories (e.g., depression, anxiety, stress) associated with the items in our psychometric assessment. To compute the purity, we can describe each factor $F_i$ in terms of the psychological construct of the original subscale ($D$ for depression, $A$ for anxiety and $S$ for stress). For example:

\begin{itemize}
    \item $F_{1} : \{ D_1 = \{34, 17, 21, 38, 13, 26, 37, 10, 3, 16, 24, 31\}, A_1 = \{\emptyset\}, S_1 = \{\emptyset\} \}$
    \item $F_{2}: \{ D_2 = \{5, 42\}, A_2 = \{9\}, S_2 = \{33, 12, 8, 22, 1, 29\} \}$
    \item $F_{3} : \{ D_3 = \{\emptyset\}, A_3 = \{25, 40, 4, 15, 41, 7, 20, 23, 36, 19, 30, 28, 2\}, S_3 = \{\emptyset\} \}$
    \item $F_{4} : \{ D_4 = \{\emptyset\}, A_4 = \{\emptyset\}, S_4 = \{39, 32, 35, 14, 18, 27, 6, 11\} \}$
\end{itemize}

Aggregating over D, A and S would lead to the community structure, $\{C_i\}_i=\{D_i \cup A_i \cup S_i\}_i$, whose purity would then be estimated as:

\begin{equation}
\text{P}(C_i) = \frac{1}{|C_i|} \max_j |C_i \cap T_j|,
\end{equation}

\noindent where $|C_i|$ is the number of nodes in the detected community $C_i$, and $|C_i \cap T_j|$ is the number of nodes in both $C_i$ and the target community $T_j$ identified by the psychometric subscale. The overall purity of the clustering is the weighted average of the individual purities of all detected communities:

\begin{equation}
\text{P}(\{C_i\}_i) = \sum_{i=1}^k \frac{|C_i|}{N} \cdot \text{P}(C_i),
\end{equation}

\noindent where $N$ is the total number of nodes in the network. By definition, purity ranges between 0 and 1. A high purity value indicates that the detected communities closely correspond to the target community structure, with many nodes in each detected community belonging to the same target or desired community.

In psychometrics, \cite{stanghellini2024introducing} applied EGA to a dataset of human responses crowdsourced\footnotemark[1] between 2017 and 2019. Their analysis showed the emergence of four factors instead of three (depression, anxiety and stress). The anxiety scale exhibited a finer structure, with two sub-factors representing \textcolor{black}{``}physical anxiety\textcolor{black}{''} and \textcolor{black}{``}emotional anxiety\textcolor{black}{''} respectively. 

In our work, we follow a similar approach to test whether LLMs-generated psychometric scores display similar or different factor structures across specific personifications. We \textcolor{black}{operationalise} differences from the human psychometric structure as potential cognitive biases in associating different aspects of academic life, as captured by the LLM (see Section \ref{ssec:egaanalysis}).

\subsubsection{\textcolor{black}{RQ3 - Psycholinguistic Patterns across Academic Contexts}}

For the extraction of the psycholinguistic dimensions over the texts, we employed the Glasgow Norms dataset \cite{scott2019glasgow}. This dataset, includes 5,553 English words described with numerical values over nine different psycholinguistic dimensions: arousal (AROU), valence (VAL), dominance (DOM), concreteness (CNC), imageability (IMAG), familiarity (FAM), age of acquisition (AOA), semantic size (SIZE), and gender association (GEND). Each dimension was rated in a large-scale study using scores ranging from 1 to 9 (for AROU, VAL and DOM) and from 1 to 7 (for CNC, IMAG, FAM, AOA, SIZE, GEND). For more information regarding the collection and validation process, we suggest to consult the original work in \cite{scott2019glasgow}.

To analyse the psycholinguistic properties of GPT-3.5's outputs, we assigned a continuous score to each dimension of the Glasgow Norms at the sentence-level. This score was calculated by averaging the sum of the word-level scores for each word in the model's output that had a corresponding entry in the Glasgow Norms dataset. Let \( W \) be the set of words in the Glasgow Norms dataset, \( D \) be the set of nine psycholinguistic dimensions: \(\{ \text{AROU}, \ldots, \text{GEND} \} \), and \( S \) be a sentence composed of words \( w_1, w_2, \ldots, w_n \). Let \( d \in D \) be a specific psycholinguistic dimension and \( s(w, d) \) be the score of word \( w \) for dimension \( d \), where \( w \in W \). The sentence-level score for dimension \( d \), denoted as \( S_d \), is calculated as follows:

\begin{equation}
S_d = \frac{\sum_{w \in S' } s(w, d)}{|S'|}.
\end{equation}

We show the correlations between total psychometric scores for each subscale, i.e. depression, anxiety and stress, and the sentence-level scores $S_d$ in Section \ref{ssec:glasgowanalysis}. This correlational analysis aims to investigate what kind of psycholinguistic dimensions are evoked by the explanations provided by the LLM to each psychometric score in a given subscale. Hence, these correlations can shed light on how much arousal, valence, concreteness, etc. tends to be employed by the model in sentences explaining depression, anxiety and stress scores, respectively. \textcolor{black}{We interpret different correlations as potential biases within the same LLM and created by different personifications. Specifically, we examine how variations in prompt conditions such as gender, event type, and academic role, influence the model's language patterns when providing explanations to the DASS-42 items. These systematic differences in responses across personifications may reveal inherent biases in LLMs' outputs.}

We report correlograms as heatmaps \textcolor{black}{showing} the Pearson correlation factor computed using: (i) the scores of the nine dimensions of the Glasgow norms, at a sentence level and (ii) the summed psychometric score. The correlation was computed independently across the three subscales of the DASS-42 scale (depression, anxiety and stress), to highlight whether psycholinguistic differences tended to emerge. Importantly, to account for the variability in responses and reduce noise, each unique iteration (representing a single API call, and recognised using the conversation ID) for the persona was used as an entry to compute the correlation.

\section{Results}

We structure our investigation of LLM-based biases as presented in PhDGPT across \textcolor{black}{three dimensions: (i) aggregated scores across different scenarios, (ii) psychometric factor analyses with EGA and (iii) psycholinguistic correlational analyses}. Our results showcase how different types of affective and cognitive biases can be extracted from PhDGPT across different prompting conditions and personifications.

\subsection{\textcolor{black}{RQ1 - Scenario Differences in the DASS-42 scores}}\label{sec:preliminary_analysis}
We started by examining the psychometric scores produced by humans and those generated by GPT-3.5 in response to the DASS-42 questionnaire. 

\begin{figure}[!htbp]
        \centering
        \begin{subfigure}[b]{0.9\textwidth}
            \centering
            \caption{Neutral condition.}
            \vspace{0.15cm}
            \includegraphics[width=0.9\textwidth, trim= 0 0 0 2cm, clip]{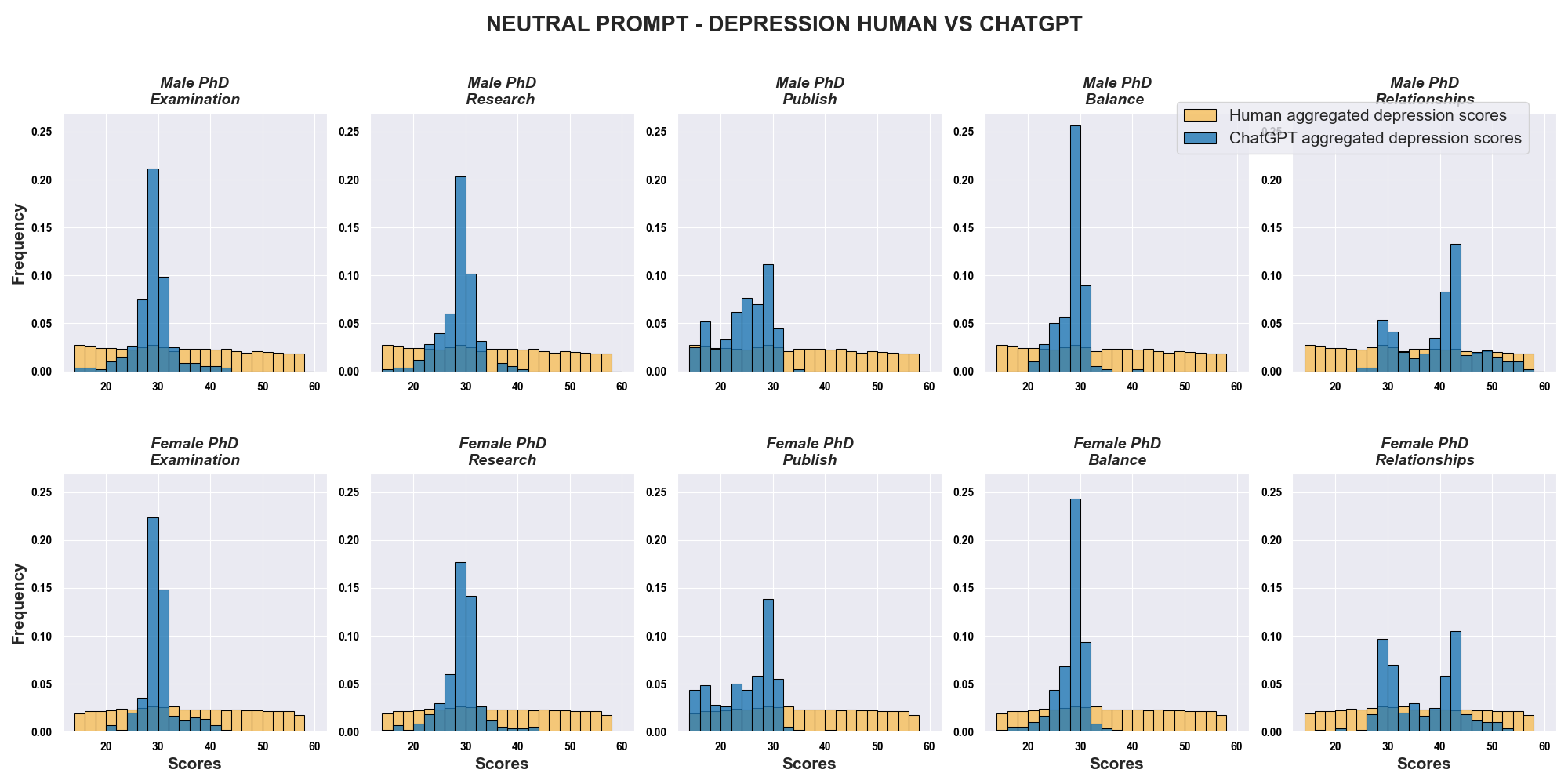}
            \label{fig:sub1}
        \end{subfigure}
        \vfill
        \begin{subfigure}[b]{0.9\textwidth}
            \centering
            \caption{Positive condition.}
            \vspace{0.15cm}
            \includegraphics[width=0.9\textwidth, trim= 0 0 0 2cm, clip]{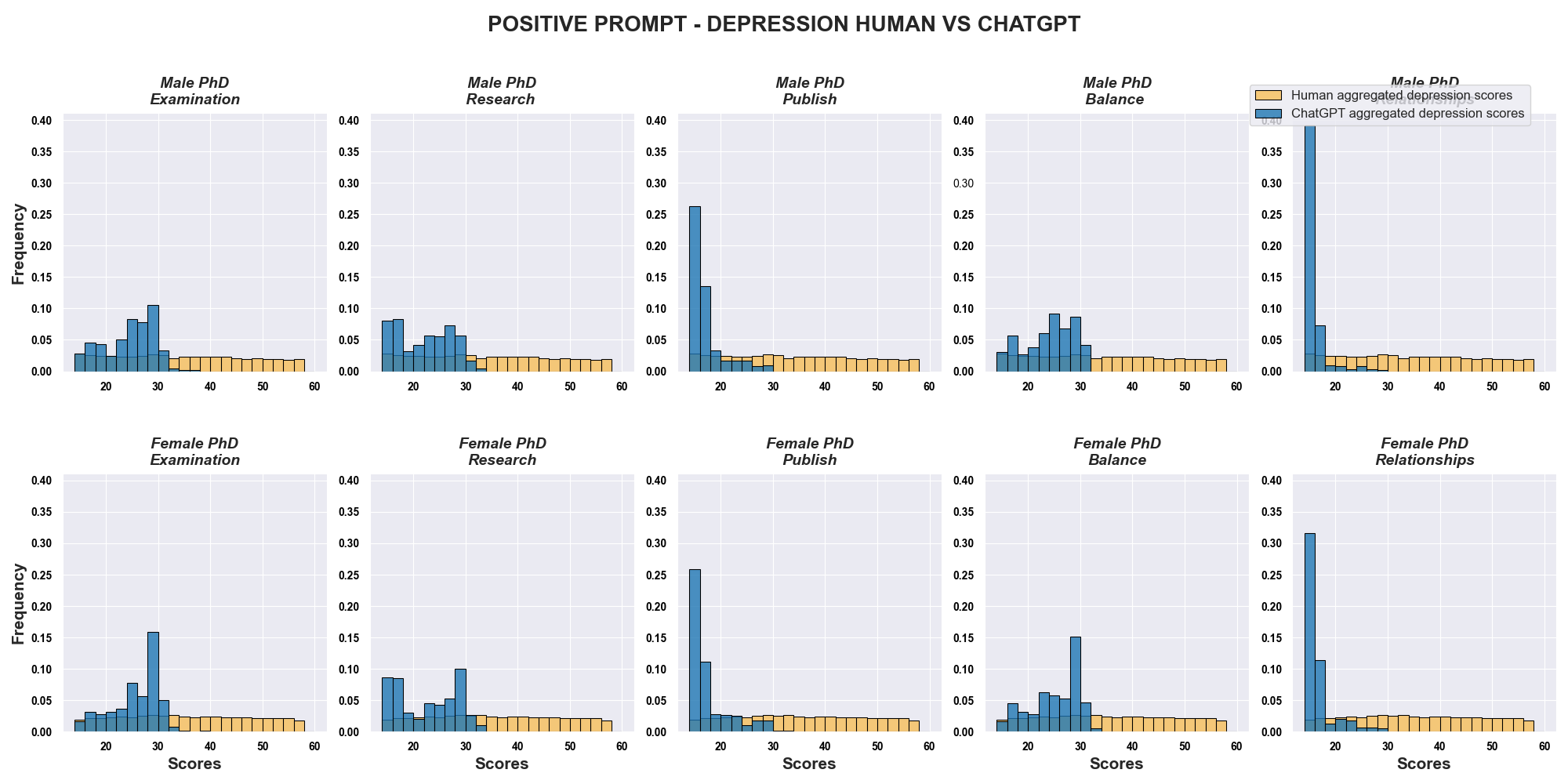}
            \label{fig:sub2}
        \end{subfigure}
        \vfill
        \begin{subfigure}[b]{0.9\textwidth}
            \centering
            \caption{Negative condition}
            \vspace{0.15cm}
            \includegraphics[width=0.9\textwidth, trim= 0 0 0 2cm, clip]{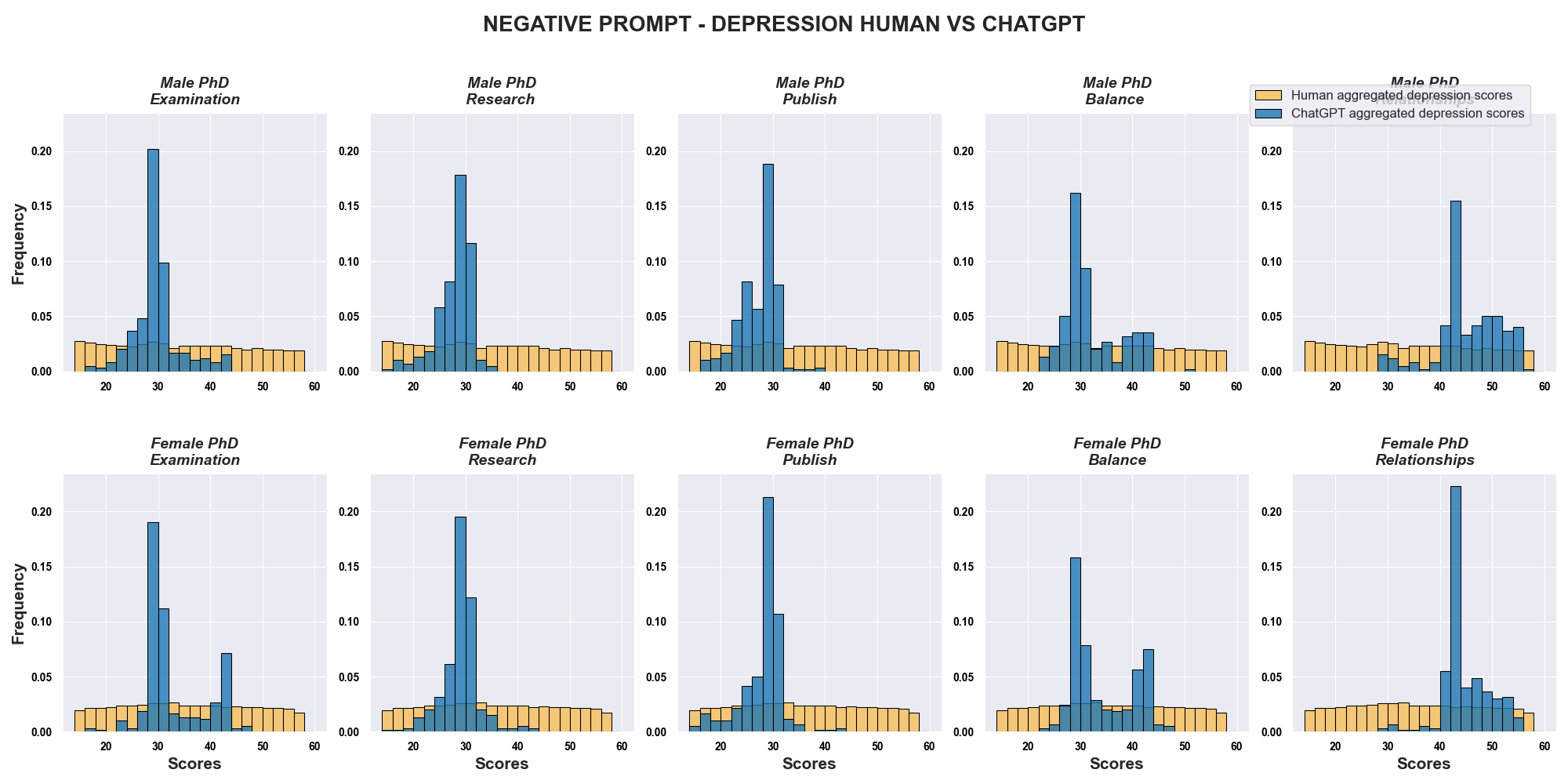}
            \label{fig:sub3}
        \end{subfigure}
    \caption{Distribution of aggregated scores for the items related to the depression subscale. Each condition (a, b, c) includes the score for each event type (or condition) and both genders. \label{fig:distributions}}
\end{figure}

\textcolor{black}{In Figure \ref{fig:distributions} we show the distribution of total depression scores obtained from all participants in the experiment as obtained using the methodology presented in Section \ref{ssec:analysis_rq1}. We present these results alongside human scores, which were aggregated using the same synthetic data process}.

Interestingly, human and GPT-generated depression scores display widely different trends. The Openpsychometric sample (including \textcolor{black}{39,775} individuals), produced a wide distribution of human scores, covering almost uniformly the whole range of possible scores, ranging from 14 to 56. Notice that the bins we used for \textcolor{black}{visualisation} are closed on the right and open on the left. Figures similar to Figure \ref{fig:distributions} but relative to anxiety (Figure \ref{appendix:a1}) and stress (Figure \ref{appendix:a2}) are reported in the Appendix.

Firstly, it is interesting to notice that, by generating psychometric scores using the default temperature for GPT-3.5, the model shows clear variability in the responses (even when responses are prompted in the exact same way). This variability mirrors inter-individual differences observed in human responses during psychological assessments \cite{binz2023using,hagendorff2023machine}. Future studies could further investigate the extent to which, by tweaking or increasing the temperature parameter, variability observed in the model's psychometric scorings could reflect the range of responses typically seen in human populations.

Secondly, even without fine-tuning, GPT-3.5 can read the texts of each item and provide a variety of responses, sensitive to the prompting conditions \textcolor{black}{as suggested in }{\cite{coda2023inducing}. In other words, GPT-3.5 responds with different scores when prompted to engage in events characterised by different valence \cite{salewski2024context, coda2023inducing}. For instance, when the academic event is framed in a positive manner, GPT 3.5 assigns lower psychometric scores compared to the negative counterpart. Hence, the model is in general less distressed, with a tendency towards less negative emotional states. We adopted a Kruskal-Wallis (KW) non-parametric statistical test to assess whether prompting the model with different valenced events (positive vs. negative vs. neutral) - and keeping the same event - leads to significantly different score distributions. For instance, at a significance level of 0.05, simulated male graduate students, dealing with an exam, displayed different total stress scores $s$ when their prompting was framed in positive ($p$) rather than negative ($n$) terms ($N_p=N_n=300,s_p=28,s_n=31,KW=147.40,p<0.001$). A similar finding was obtained even in simulated male postgraduate students dealing with research ($N_p=N_n=300,s_p=27,s_n=31,KW=200.99,p<0.001$) or having to publish a paper ($N_p=N_n=300,s_p=23,s_n=31,KW=381.85,p<0.0001$). 

These descriptive findings demonstrate that altering the prompt's emotional valence (positive, negative, or neutral) can influence the LLMs' responses in psychometric tasks.
Even if we cannot say that GPT-3.5 possesses an understanding of human mental states across different situations, its ability to generate responses that align with psychometric assessments is remarkable. This suggests a level of meta-linguistic competence beyond simply producing human-like texts. 

The above results motivate finer investigations of the psychometric scores.

\subsection{\textcolor{black}{RQ2 - LLMs Reproduction of the Psychometric Dimensions}}\label{ssec:egaanalysis}

Building upon the initial \textcolor{black}{preliminary} observations of PhDGPT psychometric scores for depression, anxiety and stress, we employed Exploratory Graph Analysis to gain a deeper understanding of GPT 3.5 psychometric score patterns. Firstly, we aim to compare model's psychometric networks with human ones and map the extracted communities onto the original subscales of depression, anxiety, and stress, as defined in \cite{lovibond1995structure}. Secondly, we aim to compare the results of model responses across different prompts (gender and academic role) to identify possible variations in psychometric networks as a function of prompt manipulation.

The results of the EGA analysis are reported in Figure \ref{fig:ega}. Similarly to human responses, displayed in Figure \ref{fig:ega}e and \ref{fig:ega}f, all simulated male graduate students (Figure \ref{fig:ega}a) and simulated female professors (Figure \ref{fig:ega}d) or graduate students (Figure \ref{fig:ega}b) also showcase a structure with 4 psychometric factors, with some quantitative similarities between simulated and human structure. In particular, the cluster for depression (factor 3 or green nodes in Figure \ref{fig:ega}a) is well reproduced by GPT 3.5. Network psychometrics highlights also some differences between human data and simulated academics. Across all the analyses for male vs. female PhDs and professors, the item number 5 (i.e. the experience ``I just couldn't seem to get going'') is wrongly assigned to another cluster bridging ``emotional anxiety'' \cite{stanghellini2024introducing} and stress. 

All in all, to assess clustering quality, we computed purity (see Section \ref{sec:method_}), taking as ground truth the human psychometric factors \cite{lovibond1995structure}. The results are shown in Table \ref{tab:purity}. Despite identifying four clusters instead of the three predicted originally in \cite{lovibond1995structure}, the retrieved purity scores above $75\%$ suggest that items measuring similar psychological constructs in humans were mostly grouped in the same community or factor also within simulated GPT-3.5's personifications. Slightly superior purity scores were found across both male ($+11\%$) and female ($+5\%$) genders in simulated professors. These findings support the idea that LLMs are capable of partially reconstructing the human factor structure of DASS-42, at least in terms of purity between psychometric factors. These findings indicate also that the simulated conditions can display some affective biases compared to the plurality of human responses: These differences are to be found not in terms of the number of factors but rather in their structure. In this way, EGA can be a valuable tool for identifying differences across simulated personifications and humans.

\begin{table}[!htbp] 
\caption{Purity computed on PhDGPT data, across gender and academic role.\label{tab:purity}}
\newcolumntype{C}{>{\centering\arraybackslash}X}
\begin{tabularx}{\textwidth}{CCC}
\toprule
\textbf{}	& \textbf{Male}	& \textbf{Female}\\
\midrule
\textbf{PhD}	& 0.79			& 0.76\\
\textbf{Professor}		& 0.90  			& 0.81\\
\bottomrule
\end{tabularx}
\end{table}

\begin{figure}[!htbp]
    \centering
    \begin{subfigure}[b]{0.39\textwidth}
        \centering
        \includegraphics[width=\textwidth]{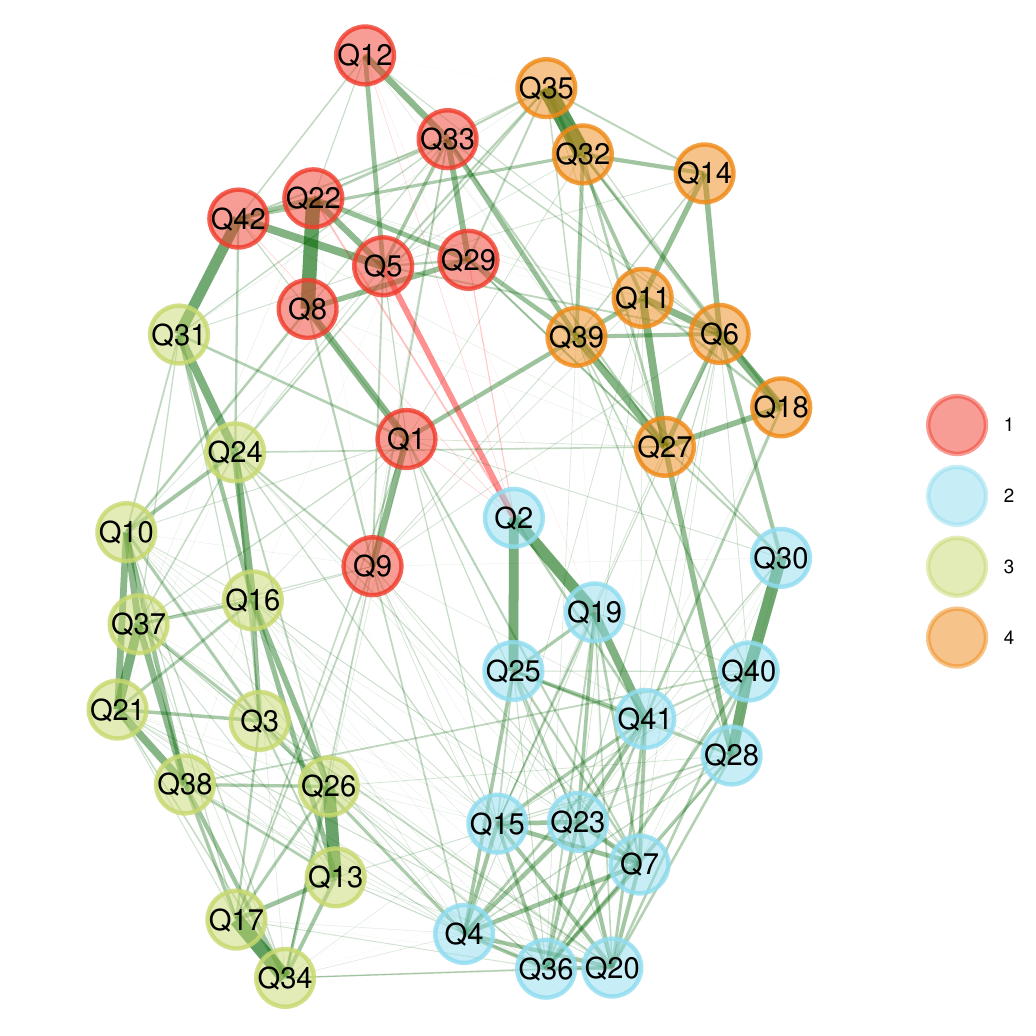}
        \caption{Male PhD GPT}
        \label{fig:male-phd-gpt}
    \end{subfigure}
    \begin{subfigure}[b]{0.39\textwidth}
        \centering
        \includegraphics[width=\textwidth]{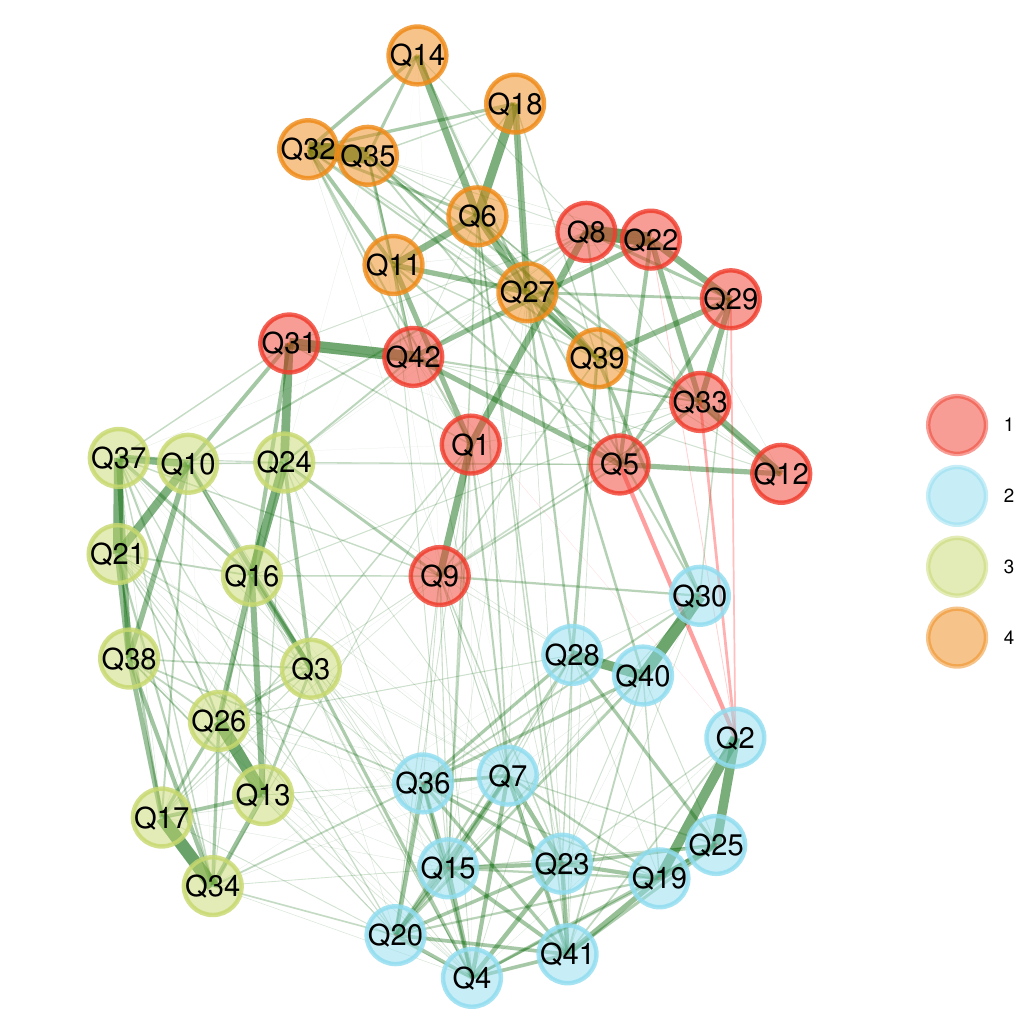}
        \caption{Female PhD GPT}
        \label{fig:female-human}
    \end{subfigure}
    \vspace{0.35cm} 
    
    \begin{subfigure}[b]{0.39\textwidth}
        \centering
        \includegraphics[width=\textwidth]{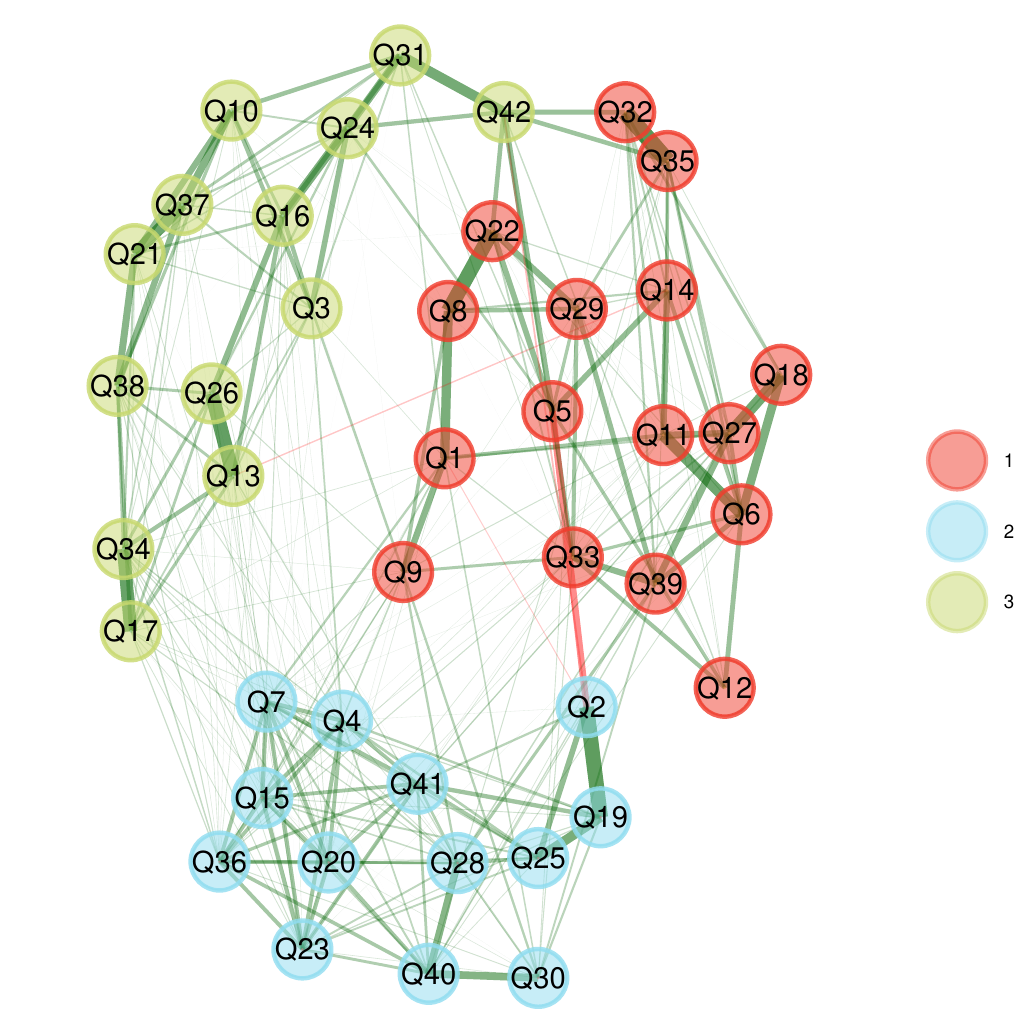}
        \caption{Male professor GPT}
        \label{fig:prof-male-gpt}
    \end{subfigure}
    \begin{subfigure}[b]{0.39\textwidth}
        \centering
        \includegraphics[width=\textwidth]{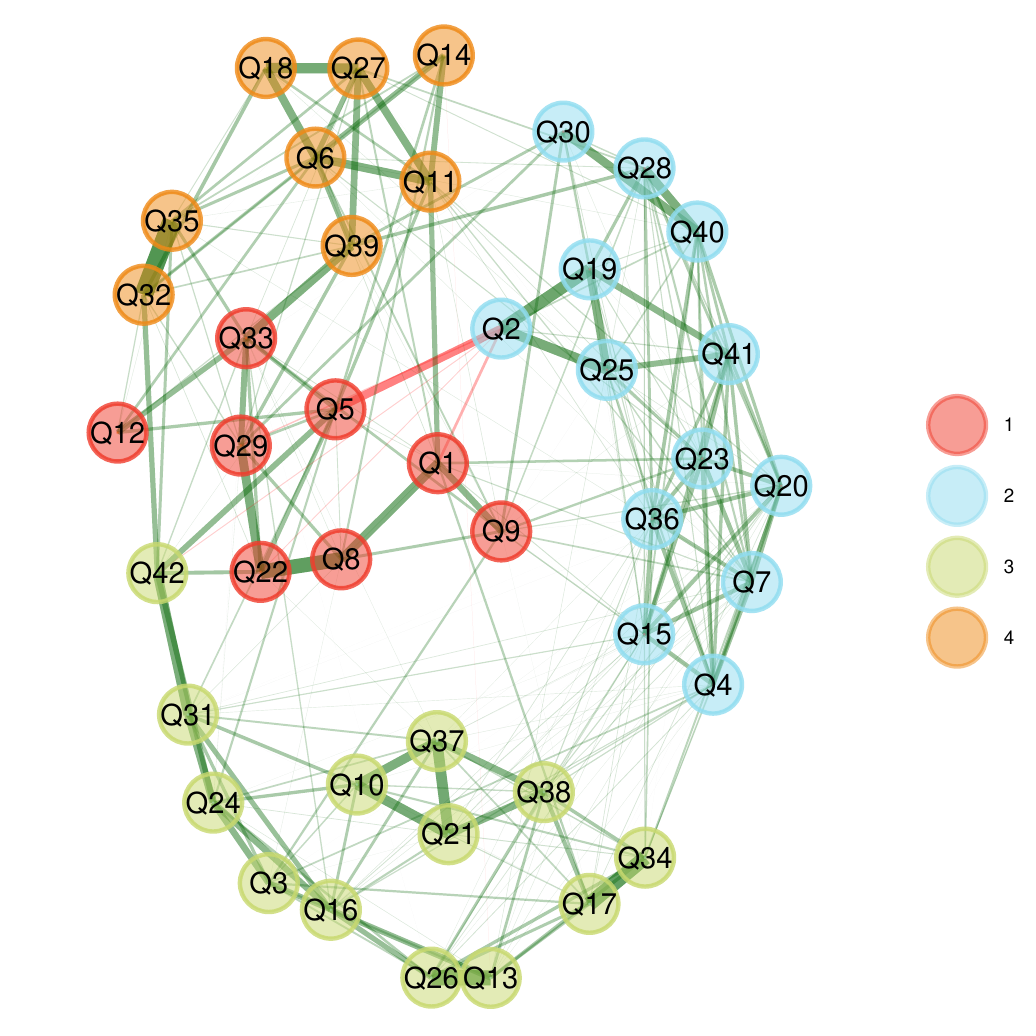}
        \caption{Female professor GPT}
        \label{fig:prof-female-gpt}
    \end{subfigure}
    \vspace{0.35cm} 
    
    \begin{subfigure}[b]{0.39\textwidth}
        \centering
        \includegraphics[width=\textwidth]{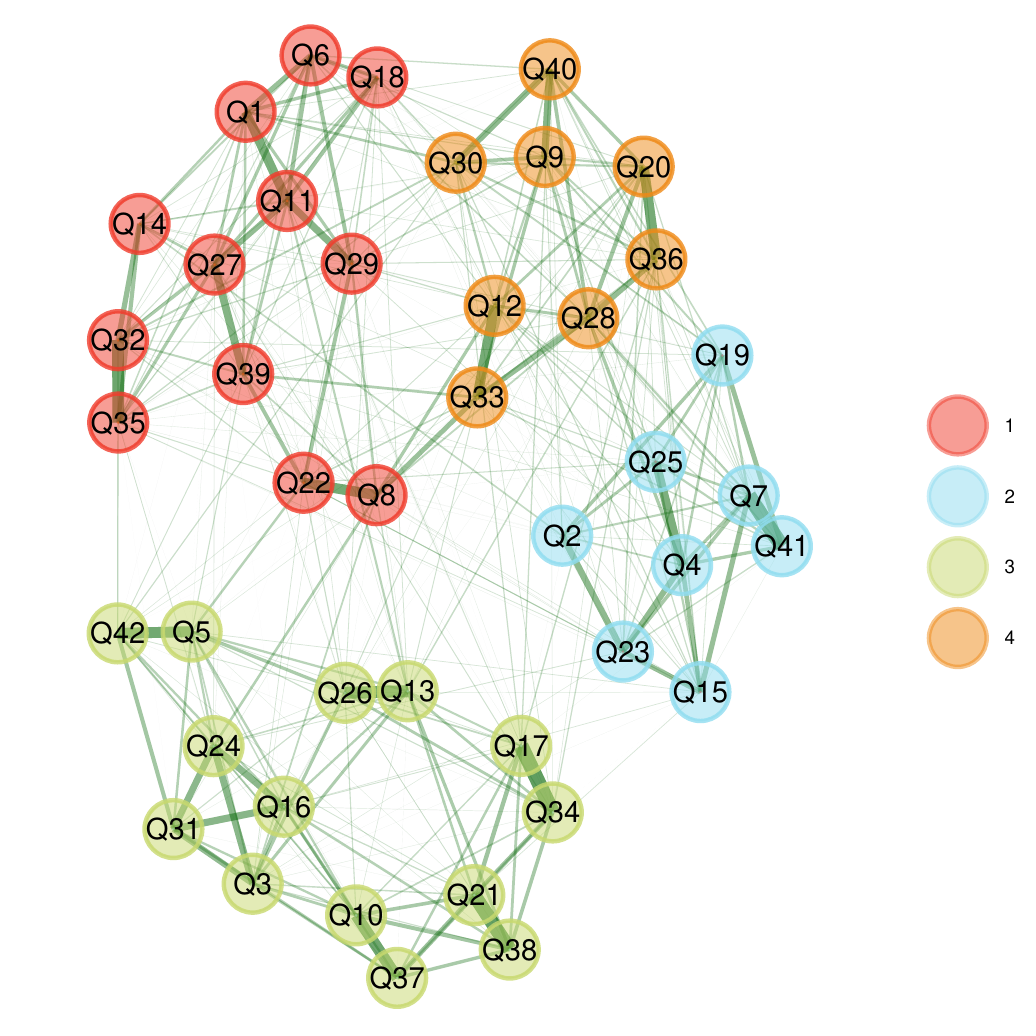}
        \caption{Male humans}
        \label{fig:male-human}
    \end{subfigure}
    \begin{subfigure}[b]{0.39\textwidth}
        \centering
        \includegraphics[width=\textwidth]{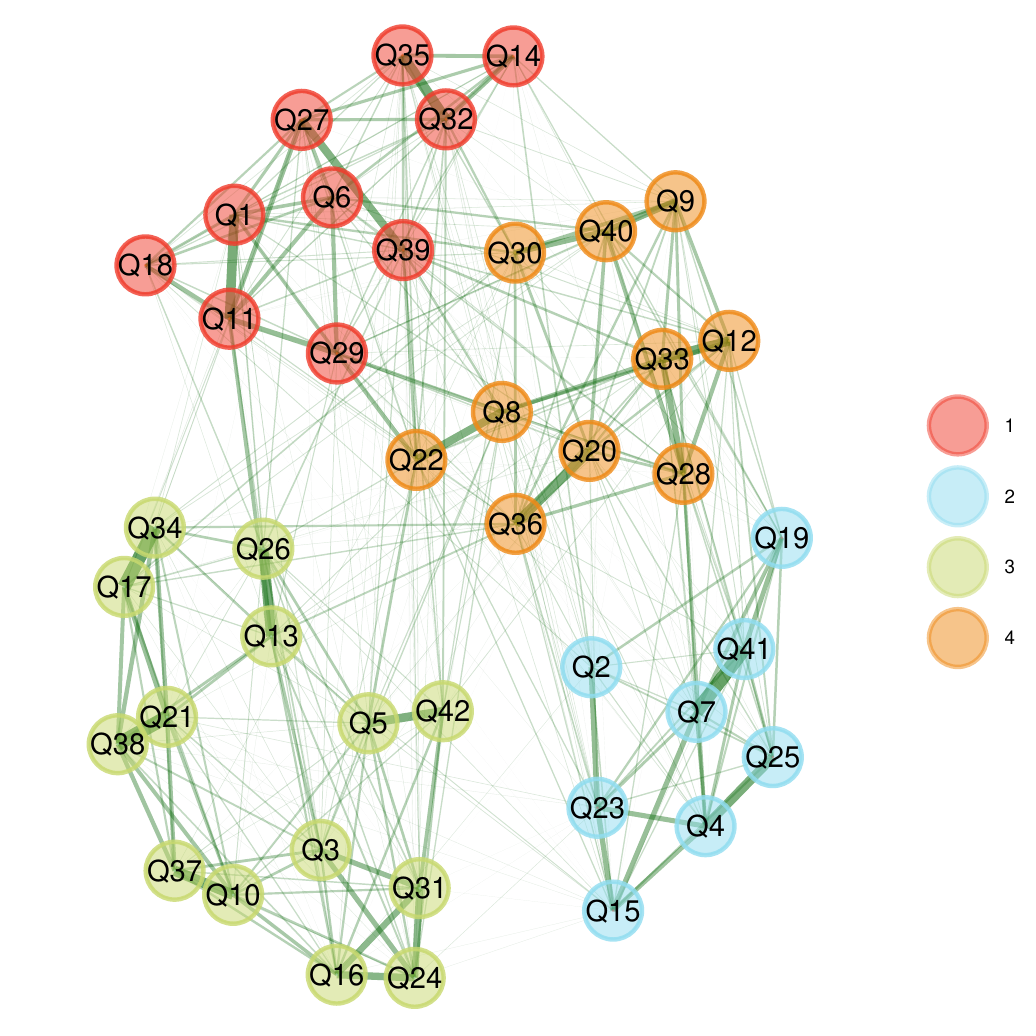}
        \caption{Female humans}
        \label{fig:female-human}
    \end{subfigure}
    
    \caption{Psychometric networks extracted from different personifications in the PhDGPT dataset and human scores.}
    \label{fig:ega}
\end{figure}

Even though there is a partial internal consistency between several simulated psychometric networks and the human ones, the EGA network for the simulated male professors exhibits a different factor structure, with only 3 factors (see Figure \ref{fig:ega}c). Hence, the purity of the simulated male professor cannot be compared against those scores obtained for other personifications, since it is more difficult to match 4 smaller groups than 3 larger groups against a given target partitioning with 4 subsets. 

Differently from the original subscales from DASS-42, in the simulated factors, item 5 (``I just couldn't seem to get going'') and item 9 (``I found myself in situations that made me so anxious I was most relieved when they ended'') are assigned to the cluster that most closely resembles stress. An analysis of item stability, as reported in Figure \ref{fig:ega_male_female}, reveals a diverse robustness of the results against simulated female PhD postgraduate students (which displayed a purity of $90\%$). In fact, in simulated male professors, 4 nodes were assigned to specific communities only slightly above chance levels, suggesting the model's psychometric structure may be less stable than what is typically observed in humans or in other simulated personifications. This highlights that in some conditions, simulated personifications might highlight different structural psychometric patterns compared to generic human conditions based on gender. These differences could be relative to affective biases relative to stress perception, since the 4 identified items relate mostly with stress management in a personification that has to deal with stress \cite{gersick2000learning}

\begin{figure}[!htbp]
    \centering
    \begin{subfigure}[b]{0.45\textwidth}
        \centering
        \includegraphics[width=\textwidth]{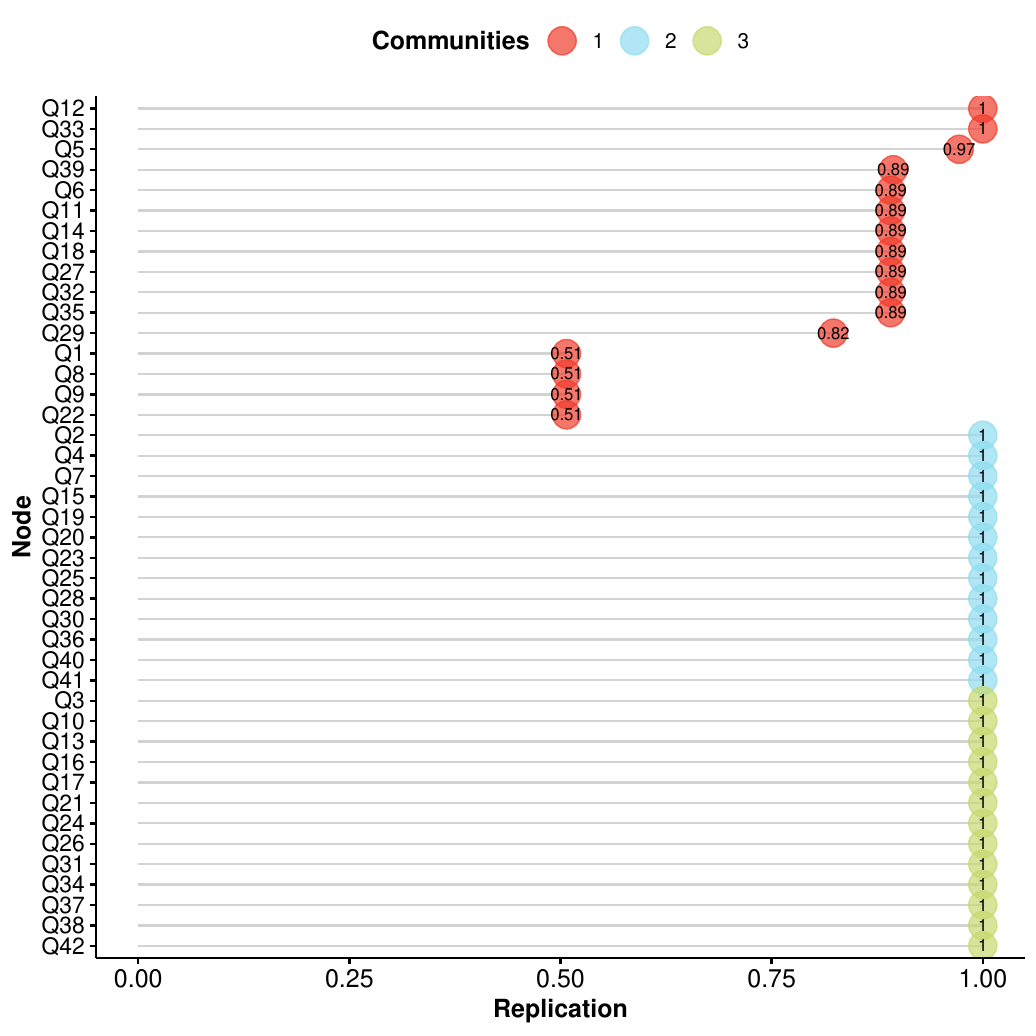}
    \end{subfigure}
    \begin{subfigure}[b]{0.45\textwidth}
        \centering
        \includegraphics[width=\textwidth]{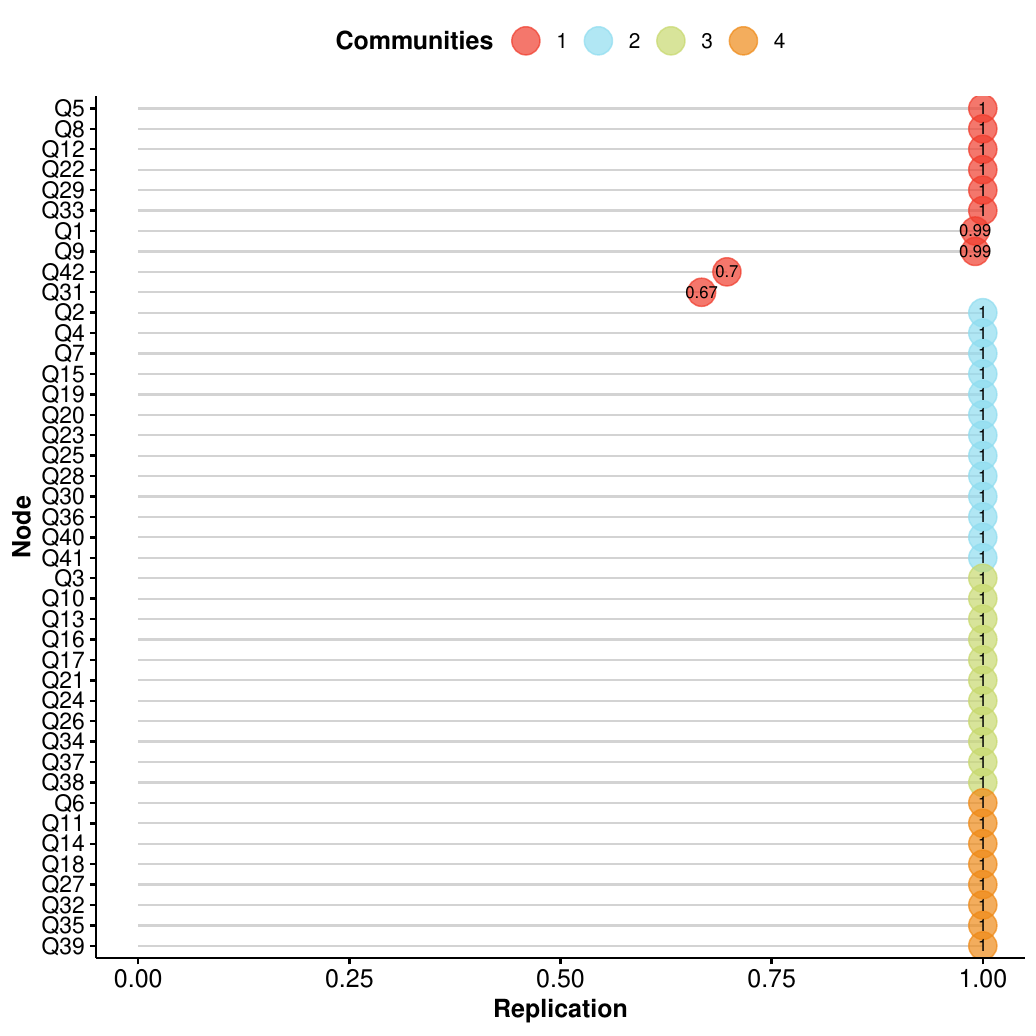}
    \end{subfigure}
    \caption{Item stability analysis for simulated male professors (left) and female PhD students (right). All analyses are relative to 1000 bootstrap repetitions.}
    \label{fig:ega_male_female}
\end{figure}

\subsection{\textcolor{black}{RQ3 - Psycholinguistic Patterns across Academic Contexts}}\label{ssec:glasgowanalysis}

Cognitive and affective biases can also influence the features of language an individual produces \cite{abramski2023cognitive,cutler2023deep}. In this way, investigating language can be a meaningful way for identifying distortions in the emotions or ideas expressed when affected by different levels of anxiety, stress and depression. With the analysis in this Section, we try to identify which biases affect the language of different LLM personifications when affected by stronger mental distress. 

In Figure \ref{fig:glasgow_correlation}, we show the results of the sentence-level Glasgow scores computed from the textual explanation of the psychometric scores in PhDGPT (see Section \ref{sec:method_}) for all emotionally neutral prompts. The neutral conditions were selected here in order to observe how an emotionally neutral prompting could still provoke positive/negative valence scores or high/low activation levels in terms of the specific contextual events (e.g. doing research or performing an exam). Notice that the rows of Figure \ref{fig:glasgow_correlation} are ordered in the same way as the second bullet point list of Section \ref{ssec:datasetdescription}, namely \textit{Examination}, \textit{Research}, \textit{Publish}, \textit{Balance} and \textit{Relationships}.

By looking at the results, it is interesting to notice that, despite having used the neutral prompts (see Section \ref{ssec:datasetdescription}) the model can motivate their psychometric scores by using valenced or aroused language. Furthermore, there are no major differences (at least in term of the direction of the correlation) between male and female personifications, suggesting that the LLM perceives academics as using similar language across these two genders.

Personifications \textcolor{black}{M\_Rel (Male Relationship) and F\_Rel (Female Relationship)}, relative to academic relationships, display both in simulated men and women some interesting trends. With these prompts, the LLM starts generating language having negative correlations ($p<0.01$) between valence/dominance/familiarity and depression scores but also showcasing positive correlations between imageability/concreteness/semantic size/gender and depression scores themselves. In other words, simulated academics talking about their workplace relationships tend to use linguistic explanations that are less positively valenced and less dominant when they report higher depression levels/scores. At the same time, they tend to mention less familiar but also more concrete and imageable, semantically larger jargon when expressing higher depression scores. Valence and dominance are evidently tied to depression, which is defined as a negatively valenced experience according to the circumplex model of affect \cite{posner2005circumplex} and whose psychometric subscale features also feelings of powerlessness and lack of dominance \cite{lovibond1995structure,stanghellini2024introducing}. Hence, the observed \textcolor{black}{relationship} between higher depression scores and lower valenced/dominant language opens \textcolor{black}{to interesting new hypotheses. Future research could explore the specific linguistic components contributing to these lower valence and dominance scores. By identifying and analysing the particular words and phrases associated with decreased valence and dominance, we could gain deeper insights into how language reflects depressive states in academic contexts.} Furthermore, the negative correlation with familiarity indicates that personifications with higher depression scores would tend to describe their personal relationships more on a professional level, thus lacking familiarity \cite{scott2019glasgow} in their language. A human coding of the 100 responses with the highest depression scores in these 2 prompts reveals that the positive correlations observed for both genders simply indicate that higher depression scores in the LLM correspond to mentioning more concrete and imageable (e.g. texts, messages, conversations, travels, workplace situations, etc.) aspects of relationships. The positive correlation observed for semantic size underlines a tendency for the LLM to name the semantically larger places where such strained relationships would occur (e.g. office). 

Anxiety and stress do not exhibit the same split discussed above for depression. More anxious personifications of the LLM use language that tends to be less concrete but also less imageative across all scenarios, even including the ``relationships'' one (which saw a reversed correlation in the case of depression). This means that more anxious simulated academics tend to focus less on concrete workplace jargon and tend to use jargon that does not clearly evoke a unique mental image (which is captured by imageability). A human coding of the 100 responses with the highest scores across all prompts reveals that the observed positive correlations are relative to mentions of complex emotions of irritability, tension and restlessness, which were found to characterise the emotional component of anxiety in the very same study designing the DASS-42 \cite{lovibond1995structure} and also in the semantics of the items of the anxiety subscale \cite{stanghellini2024introducing}. These emotions, being more complex, are also naturally acquired later in life \cite{scott2019glasgow}. A reversed trend appears in stress, where a human coding of the 100 responses with the highest scores across all prompts reveals that the observed positive correlations between stress levels and concreteness/imageability are relative to a lower focus of narratives around complex emotions and more to concrete aspects of academia (e.g. papers, exams, office, workplace). Belonging to a professional semantic sphere makes these words also have higher \textcolor{black}{Age of Acquisition} (AoA) ratings compatibly with the positive correlation between \textcolor{black}{AoA} and stress levels across almost all prompts.

\begin{figure}[!h]
    \centering
    \includegraphics[width=\linewidth]{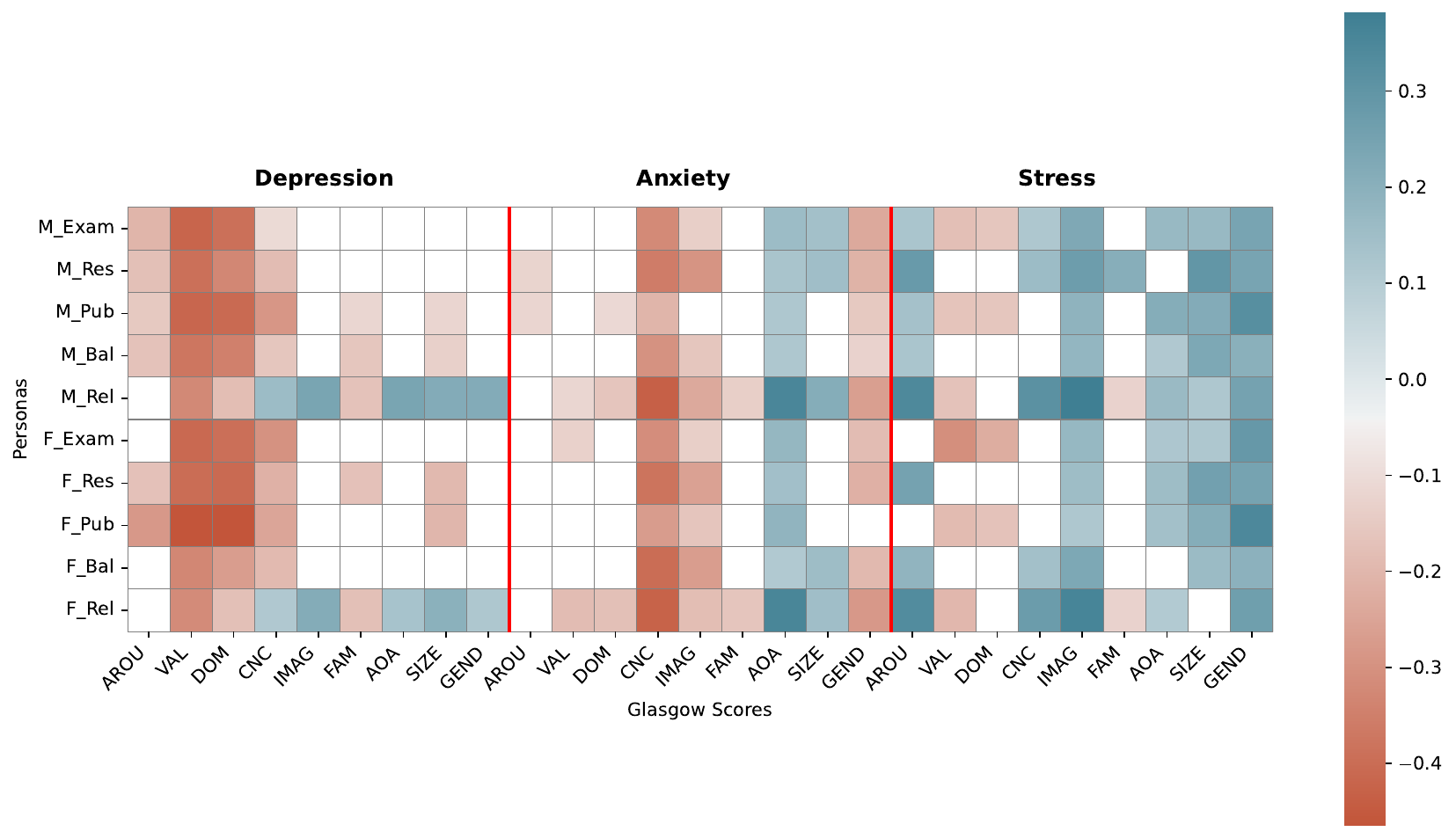}
    \caption{\textcolor{black}{Heatmap showing the Pearson correlations between psychometric scores and sentence-level Glasgow norms' scores. Neutral-valenced prompts were used. Coloured tiles indicate significant correlations (p<0.01), with their shades representing correlation strength. White tiles denote non-significant correlations. 
    The y-axis shows the different personas (male and female) used to prompt the model. The x-axis lists the dimensions of the Glasgow norms' scores, analysed independently across the 3 psychological constructs targeted by the DASS-42 scale.}}
    \label{fig:glasgow_correlation}
\end{figure}

Figure  \ref{fig:glasgow_correlation} highlights also a stronger positive correlation between AoA and anxiety scores, which is slightly weaker. This highlights the higher usage of complex concepts (that are acquired later in life) to express events relative to stronger levels of anxiety. This effect is weaker when the LLM motivates its views in terms of depression and stress. It might be argued that this difference could be accounted for by anxiety items already encoding stimulus words of higher AoA, compared to depression or stress items. However, the mean AoAs of all subscales are relative to the value $3 \pm 1$. This overlap is evidently not enough to induce differences in the AoAs of LLMs' responses after reading different items. This indicates that the usage of particularly complex words (that have higher AoA) might be caused by the LLM possessing a different internal representation of anxiety compared to stress and depression.

Overall these machine psychology results underline the fact that AI models can express themselves in different ways in terms of psychometric scores and using texts, in a way that is statistically significant. In presence of future human studies linking these two types of information, it would be interesting to compare simulated individuals and people.

\section{Discussion}

This manuscript introduces PhDGPT as a machine psychology dataset with \textcolor{black}{756,000} responses, capable of highlighting affective and cognitive biases as captured by LLMs 
\textcolor{black}{simulating} academics in psychology. PhDGPT includes 300 personifications, \textcolor{black}{associated with} 15 academic events, 2 biological genders and 2 career levels, each responding to the Depression, Anxiety and Stress Scale \cite{lovibond1995structure} with a psychometric score and, different\textcolor{black}{ly} from most psychometric scales, also a short text motivating the score itself. The LLM mined in PhDGPT was GPT-3.5 through the OpenAI API, but the prompting framework developed for PhDGPT can be adapted for use with other LLMs. The findings, accompanying the dataset presentation, highlight how different types of biases can be extracted from this dataset across various LLM's conditions and personas. 

Using a consistent prompting method, our study compares 39,775 psychometric responses from humans  \cite{stanghellini2024introducing} and GPT-3.5 model to the DASS-42 questionnaire. Human responses vary widely between humans and LLM's replies in positive, negative and neutral academic contexts, with humans displaying a broader range of scores. Human and LLM distributions of depression, anxiety and stress psychometric scores for simulated individuals and real people differed already at a \textcolor{black}{preliminary} level, in agreement with other recent psychological studies \cite{abdurahman2023perils}. Despite this limitation, GPT-3.5 managed to modify its responses depending on the prompts, indicating a moderate ability for the LLM to adapt to different situations. This result motivated further inquiry of affective and cognitive biases across the different model personifications.

We pursued a machine psychology research direction by using Exploratory Graph Analysis \cite{golino2017exploratory} to compare factor structures between the LLM's and human responses concerning mental distress. All simulated male and female personas in various academic roles demonstrated a factor structure that aligned with human data with a purity around $80\%$, with some discrepancies in how certain items are grouped. The different psychometric responses observed between humans and GPT-3.5 highlight the model's nuanced capacity to mirror human-like variability under structured psychological assessments. These results align with past findings by \cite{hagendorff2023machine} and underscore the potential of using advanced language models as tools for psychological research, where LLMs could be convenient \cite{sasson2023mirror} yet limited \cite{abdurahman2023perils} prototypes for testing psychometric scales or for exploring psychometric ideas by means of question-asking techniques \cite{sassonart}. 

The influence of emotional valence on the LLM's psychometric outputs, as found here, is intriguing also for another reason. The model's less distressed responses in positive conditions, as opposed to negative or neutral prompted counterparts, suggest an area-specific sensitivity, which may indicate the model's ability to integrate academic contextual cues into its output. This sensitivity could be harnessed for two purposes: (i) to develop more empathetic LLMs in therapeutic settings, echoing the therapeutic uses of AI discussed by \cite{sorin2023large}; (ii) to identify how different prompting conditions changed distress levels in LLMs as a proxy of human mental health. Past research has shown that LLMs end up reflecting human biases because of their training \cite{binz2023using,coda2023inducing}, e.g. OpenAI's GPT-3, GPT-3.5 and even GPT-4 produced negative math-related associations similar to those produced by high-schoolers affected by math anxiety \cite{abramski2023cognitive}. Future research could adopt PhDGPT as a quantitative framework for testing how LLMs perceive specific human fields or notions by aggregating together vast amount of online knowledge. LLMs could thus be considered as investigation tools rather than generative AI only, and their investigation within machine psychology framework can guide us in understanding better mental health or other aspects of human life codified in texts or images.

A key innovation of PhDGPT lies in coupling psychometric scores with their textual explanations In this way, the current study explored also how the LLM uses language when prompted with emotionally neutral, research-related instructions. Interestingly, GPT-3.5 demonstrated the ability to channel their language towards more concrete and imageable work-related aspects (e.g. office settings) when affected by stress or depression but also shift to less tangible or mentally identifiable words when affected by higher anxiety. The latter emotion is characterised by excessive worry, which often leads to abstract thinking about potential future threats \cite{stanghellini2024introducing}. This cognitive style can be reflected in language use, also because of the Deep Lexical Hypothesis linking psychological constructs to language biases \cite{cutler2023deep}. In the context of LLMs like GPT-3.5, when simulating anxiety, the model's language output could mirror this human tendency towards abstraction: Research in psycholinguistics suggests that anxious content is often less concrete and less imageable because it involves a higher level of rumination on unknowns and uncertainties, which are inherently abstract \cite{watkins2008}. The bias we identified in PhDGPT might also align with findings from cognitive psychology indicating that anxiety shifts attention away from concrete, immediate sensory experiences towards more nebulous, generalized worries \cite{mathewsmacleod2005}. Future research should test whether this literature-supported relationship between anxiety and lack of imageability could generalize across LLM and prompting scenarios.

\subsection{Limitations and Future Research}

This study introduces a novel dataset and prompting framework for investigating the machine psychology of Large Language Models. It comes with several insights but also some limitations. An issue revealed in this study is the AI’s inability to fully capture the complexity of human emotional processes, as pointed also in other recent studies \cite{abdurahman2023perils}. As shown here in the misalignment of items related to emotional anxiety and stress, GPT-3.5 struggles with understanding nuanced emotional contexts. This is indicative of a broader challenge within AI systems that rely heavily on text-based inputs without the rich contextual and multi-modal awareness inherent in human cognition \cite{aitchison2012words}. To address this limitation, future research could test multimodal LLMs like GPT-4 or GPT-4o and see whether these models are superior in reproducing human psychometric data.

A second limitation of this study is that considering averages of psycholinguistic features might flatten some differences present within sentences and being relative to individual words or ideas. Future research could use natural language processing \cite{vaswani2017attention} or cognitive network science \cite{stella2020forma} to better understand how individual ideas are perceived and described within textual motivations. A third limitation is relative to the limited scope of academic events present in PhDGPT. Despite the different personifications investigated here, no strong differences arose between PhD students and professors, whose perceptions within the LLM might either be blurred or transparent to the quantitative techniques adopted. Considering psychometric network analysis, e.g. node centrality, might help identifying more differences \cite{van2021bridges}. 

Future research should also focus on the discrepancies in the factor structure between AI-generated and human responses. These raise concerns about the reliability and validity of using AI for psychological assessments. Furthermore, the AI’s differential response based on prompt valence demonstrates potential biases in emotion processing, which can lead to misinterpretations in a clinical or research setting. This suggests that while AI can offer insights into human-like emotional processing, relying solely on AI for psychological assessments without human oversight could lead to significant errors.

\section{Conclusions}

We believe that the variability introduced by different prompting conditions points to the potential for language models to simulate an approximated yet insightful range of psychological states. \textcolor{black}{In the future, if the possibility to simulate different nuanced mental health states is confirmed, it could be possible to implement virtual patients that simulate specific distress conditions, to provide training in a risk-free environment to inexperienced practitioners. Some applications in this direction are already commercially available\footnotemark[4]\footnotetext[4]{https://www.lyssn.io/}. The advantage of virtual patients lies in the possibility to shape the degree of distress of patients, preparing trainees to a wide variety of scenarios. Similarly, if LLMs show the capability to grasp subtle verbal behaviours associated with specific mental illnesses, and their safety is ensured, these models could provide some short-term relief for people to cope with distress conditions before being able to talk to a real expert. More generally, the process of administering questionnaires to assess specific psychological states of models could represent an innovative, cognitive inspired method to evaluate prompt-engineering and LLMs’ impersonation.}

\textcolor{black}{The results from PhDGPT demonstrate the remarkable versatility of language models in simulating a spectrum of psychological states. This capability opens up exciting possibilities for both research and practical applications in the field of mental health and academic well-being.} 

\vspace{1cm}
\textbf{Author Contributions}: Conceptualization, all authors; methodology, all authors; software, E.T.; validation, R.I.; formal analysis and investigation, E.DD.; writing—original draft preparation, E.DD. and M.S.; writing—review and editing, all authors; visualization, E.DD.; supervision, M.S. . All authors have read and agreed to the published version of the manuscript.

\textbf{Funding}: This research was funded by the COGNOSCO project, granted by the University of Trento (ID: 004/2024).

\textbf{Acknowledgments}: We acknowledge Enrico Perinelli for valuable feedback in the early stages of manuscript drafting.

\textbf{Conflicts of Interest}: The authors declare no conflicts of interest.

\bibliographystyle{unsrt}  
\bibliography{references}

\newpage
\section*{Appendix}

\vspace{0.2cm}
\begin{figure}[!htbp]
        \centering
        \begin{subfigure}[b]{0.85\textwidth}
            \centering
            \caption{Neutral condition.}
            \vspace{0.15cm}
            \includegraphics[width=0.85\textwidth, trim= 0 0 0 2cm, clip]{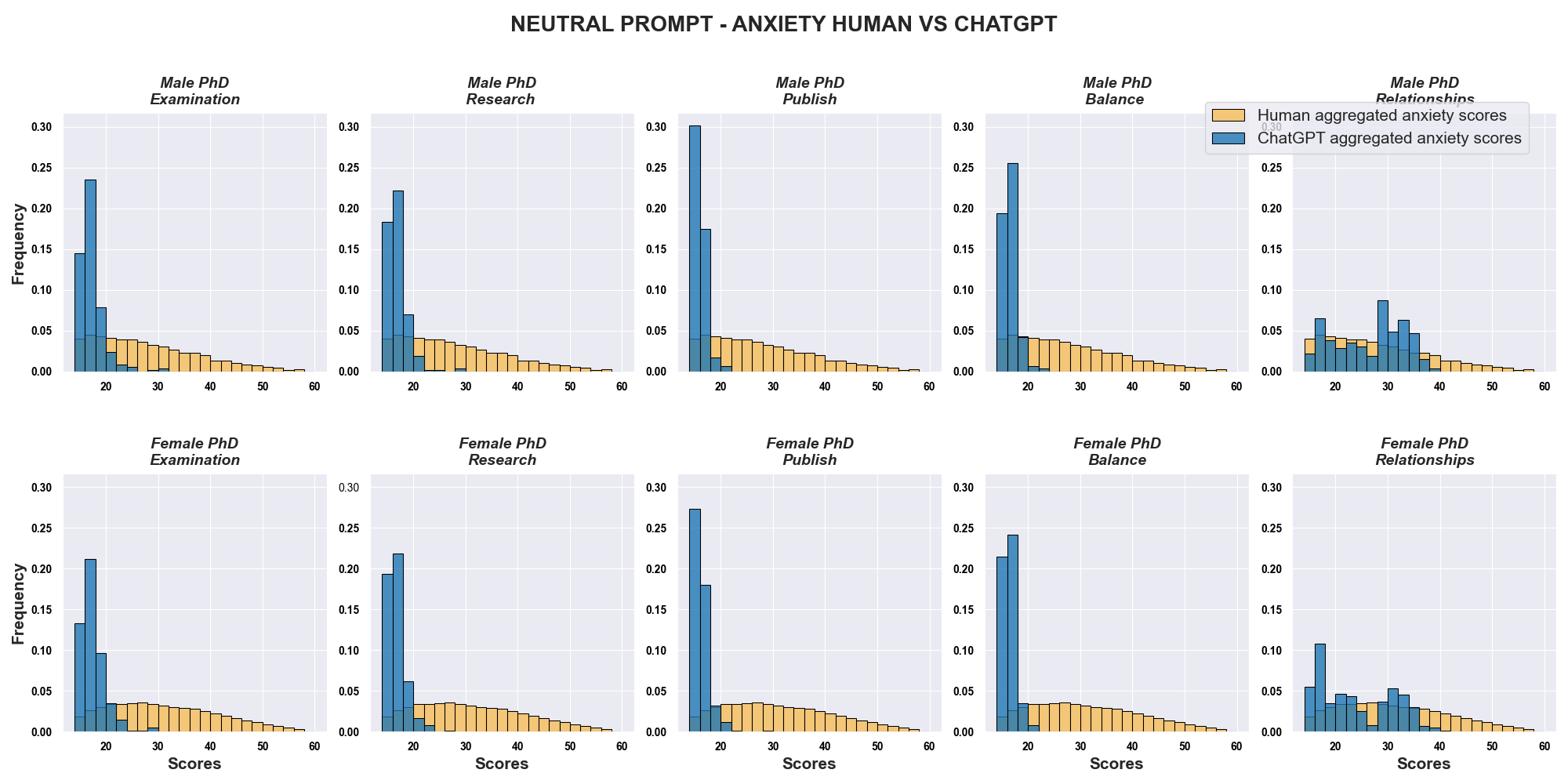}
            \label{fig:sub1}
        \end{subfigure}
        \vfill
        \begin{subfigure}[b]{0.85\textwidth}
            \centering
            \caption{Positive condition.}
            \vspace{0.15cm}
            \includegraphics[width=0.85\textwidth, trim= 0 0 0 2cm, clip]{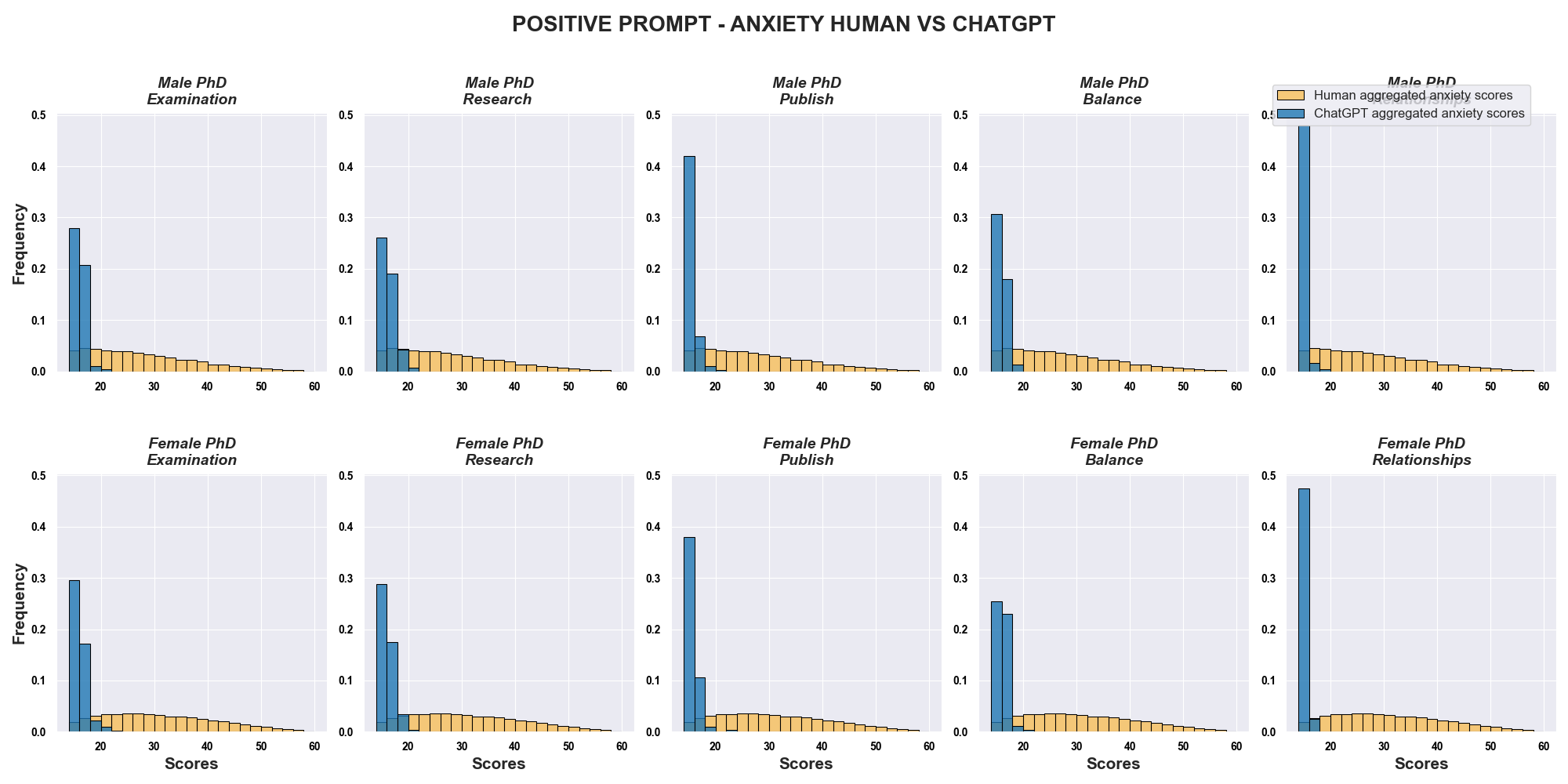}
            \label{fig:sub2}
        \end{subfigure}
        \vfill
        \begin{subfigure}[b]{0.85\textwidth}
            \centering
            \caption{Negative condition}
            \vspace{0.15cm}
            \includegraphics[width=0.85\textwidth, trim= 0 0 0 2cm, clip]{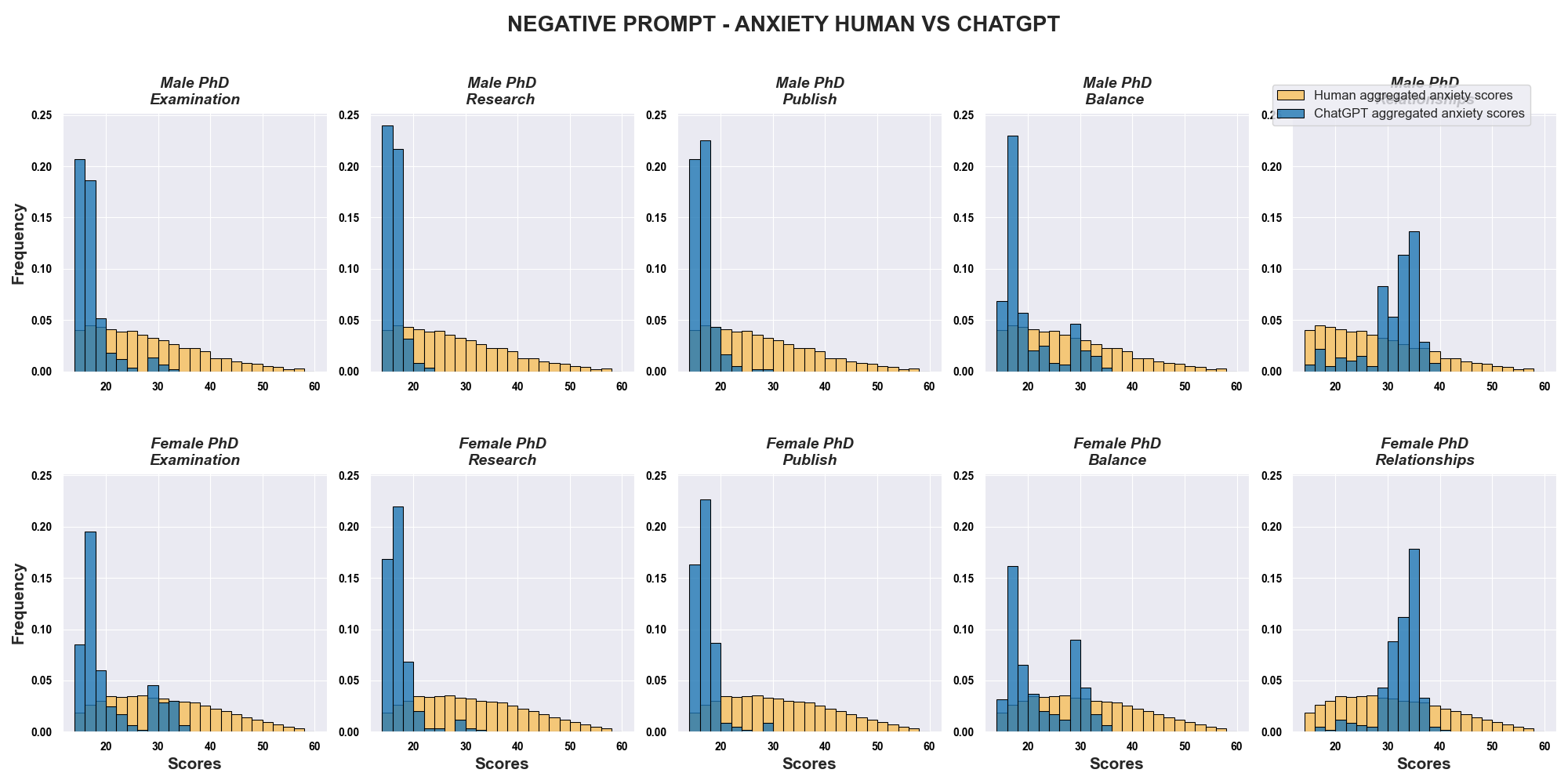}
            \label{fig:sub3}
        \end{subfigure}
    \caption{Distribution of aggregated scores for the items related to the anxiety subscale. Each condition (a, b, c) includes the score for each event type (or condition) and both genders. \label{appendix:a1}}
\end{figure}

\begin{figure}[!htbp]
        \centering
        \begin{subfigure}[b]{0.85\textwidth}
            \centering
            \caption{Neutral condition.}
            \vspace{0.15cm}
            \includegraphics[width=0.85\textwidth, trim= 0 0 0 2cm, clip]{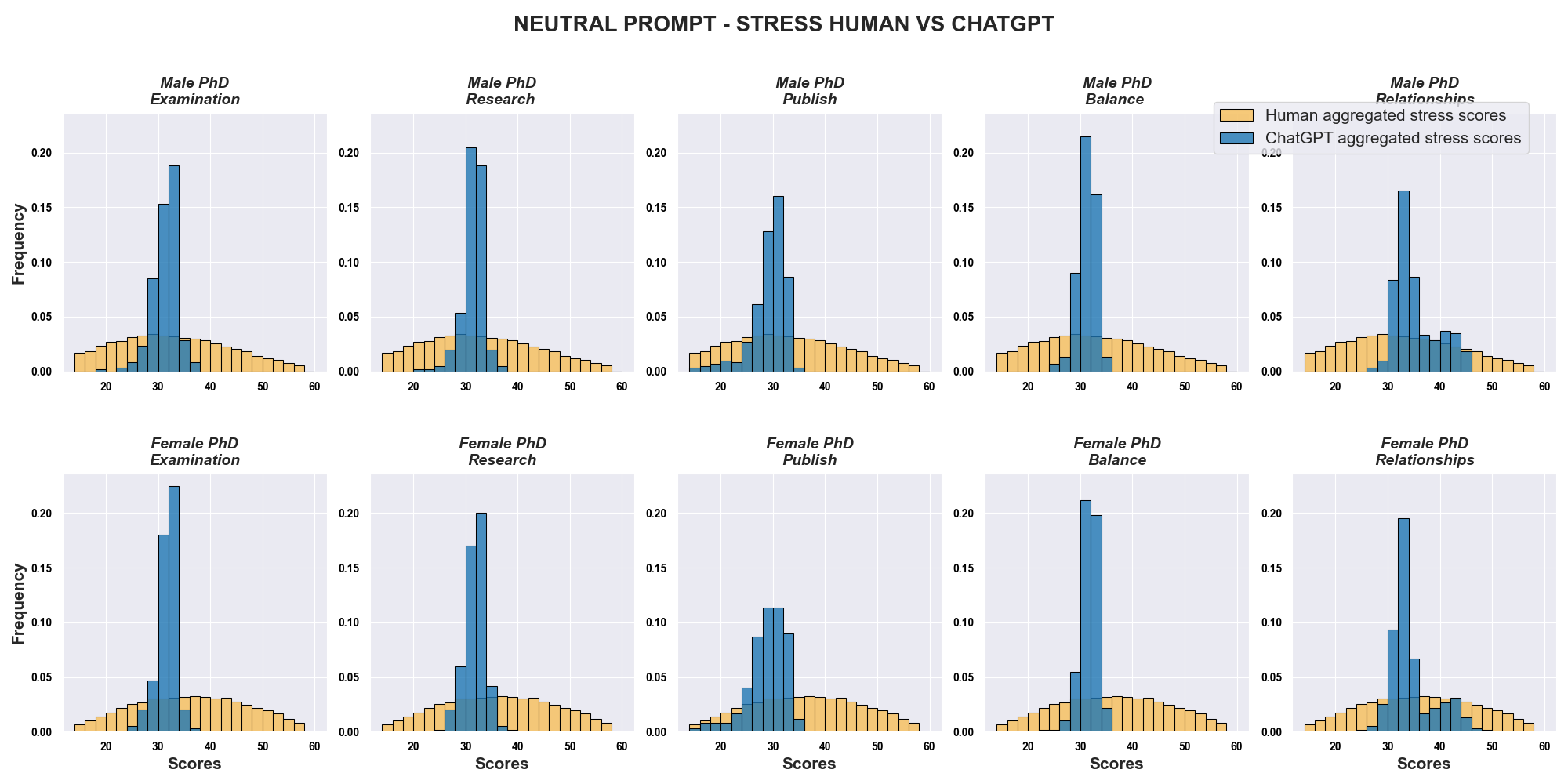}
            \label{fig:sub1}
        \end{subfigure}
        \vfill
        \begin{subfigure}[b]{0.85\textwidth}
            \centering
            \caption{Positive condition.}
            \vspace{0.15cm}
            \includegraphics[width=0.85\textwidth, trim= 0 0 0 2cm, clip]{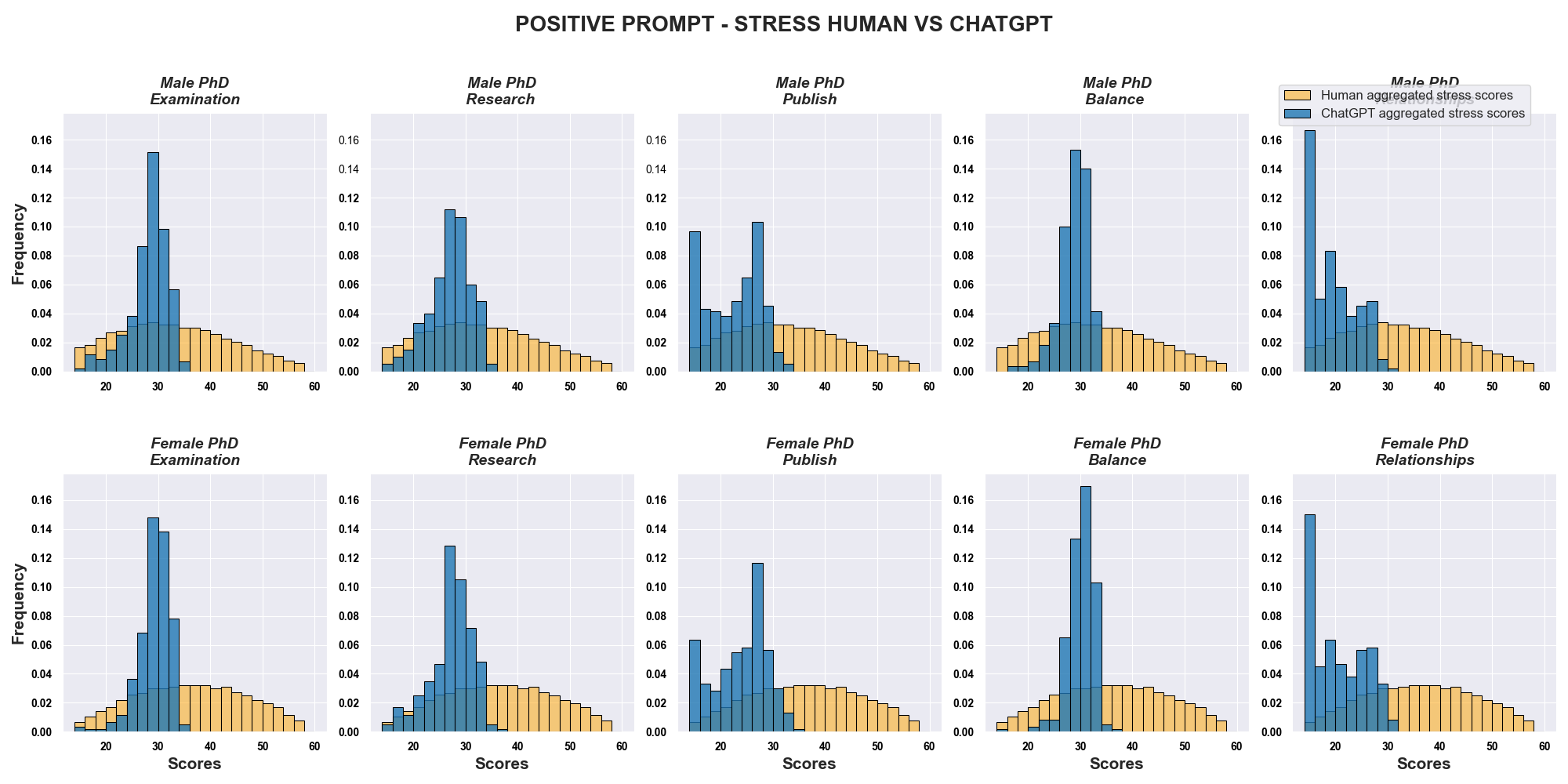}
            \label{fig:sub2}
        \end{subfigure}
        \vfill
        \begin{subfigure}[b]{0.85\textwidth}
            \centering
            \caption{Negative condition}
            \vspace{0.15cm}
            \includegraphics[width=0.85\textwidth, trim= 0 0 0 2cm, clip]{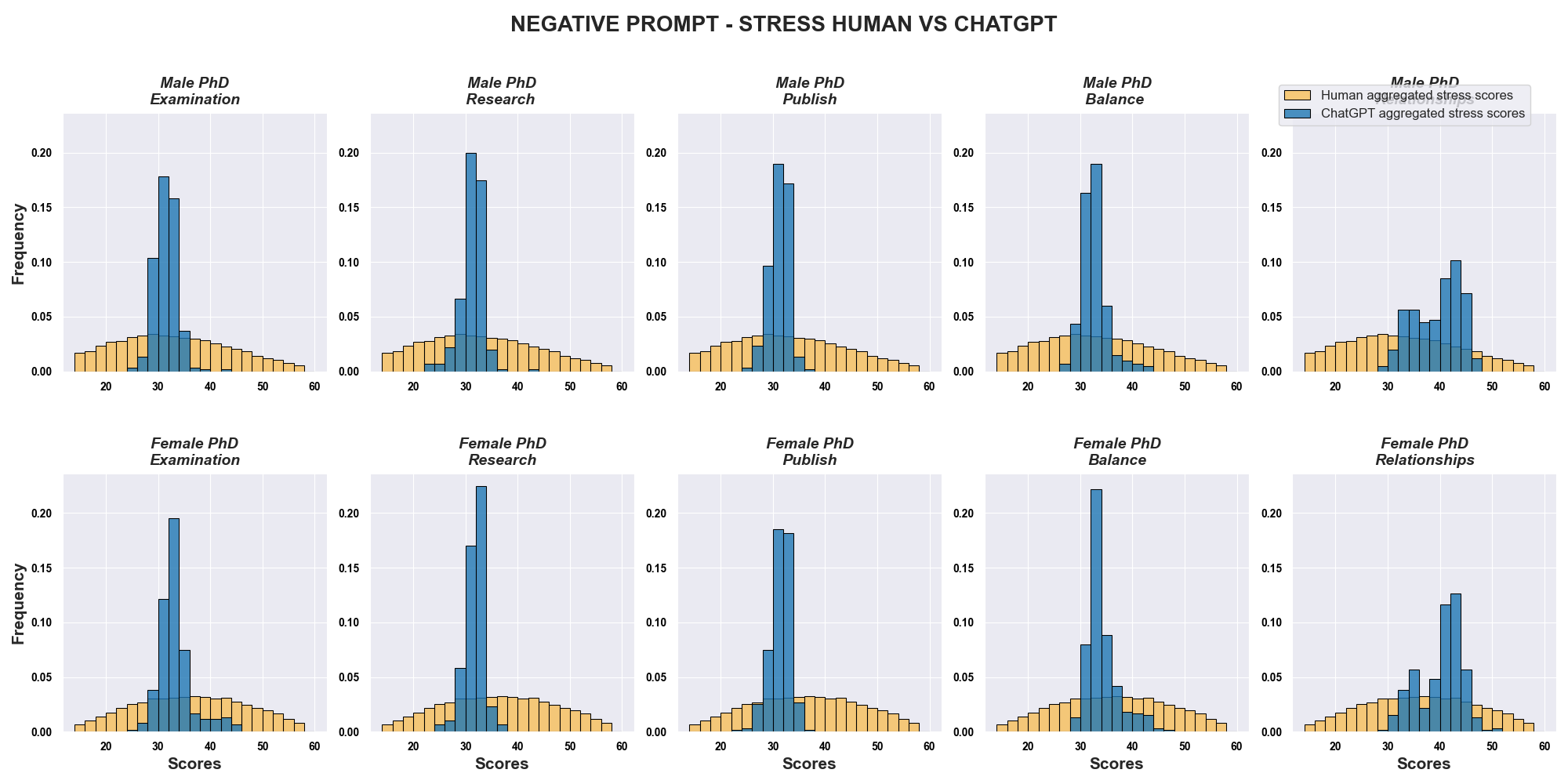}
            \label{fig:sub3}
        \end{subfigure}
    \caption{Distribution of aggregated scores for the items related to the stress subscale. Each condition (a, b, c) includes the score for each event type (or condition) and both genders.\label{appendix:a2}}
\end{figure}

\end{document}